\documentclass[journal]{IEEEtran}

\usepackage{amsfonts}
\usepackage{cite,graphicx,amsmath,amsthm,amssymb}
\usepackage{subfigure}
\usepackage{fancyhdr}
\usepackage{dsfont}
\usepackage{array,color}
\usepackage{bm}
\usepackage{float}
\usepackage{algorithm}
\usepackage{algpseudocode}
\usepackage{multirow}
\usepackage{booktabs}
\usepackage{multirow}
\usepackage{makecell}
\usepackage{colortbl}
\usepackage{graphicx}
\usepackage{stfloats}
\usepackage{subfigure} 
\usepackage{cite}
\usepackage{epstopdf}

\allowdisplaybreaks[4]

\newtheorem{theorem}{Theorem}

\newtheorem{lemma}{Lemma}

\newtheorem{proposition}{Proposition}

\newtheorem{corollary}{Corollary}

\newtheorem{property}{Property}

\newtheorem{remark}{Remark}

\newtheorem{claim}{Claim}

\makeatletter
\renewcommand{\maketag@@@}[1]{\hbox{\m@th\normalsize\normalfont#1}}%
\makeatother

\begin{document}

	\title{\huge Rate-Splitting with Hybrid Messages: DoF Analysis  of the  Two-User MIMO Broadcast Channel  with Imperfect CSIT}
	
	\author{Tong Zhang, \textit{Member IEEE},    Yufan Zhuang,
		Gaojie Chen, \textit{Senior Member IEEE}, \\Shuai Wang,  \textit{Senior Member IEEE},  Bojie Lv,      Rui Wang,  \textit{Member IEEE},  and Pei Xiao, \textit{Senior Member IEEE},   %Yinfei Xu, \textit{Member IEEE}, Gaojie Chen \textit{Senior Member IEEE} %and Tobias J. Oechtering
		
		\thanks{
			%Manuscript received 17 August 2023; revised 24 January 2024;
			%accepted 25 February 2024.  
			This work was supported in part
			by the National Key Research and Development Program of China under
			Grant 2021YFB3300200; in part by the National Natural Science Foundation
			of China under Grant 62171213, Grant 62371444; in
			part by Guangdong Basic and Applied Basic Research Project under Grant 2021B1515120067. (\textit{Corresponding author: Gaojie Chen}).

			T.~Zhang and G. Chen are with the School of Flexible Electronics (SoFE), Sun Yat-sen University, Sun Yat-sen University, Shenzhen, Guangdong 518107, China (e-mail: bennyzhangtong@yahoo.com, gaojie.chen@ieee.org).
			
			Y.~Zhuang is with the Hong Kong University of Science and Technology, Hong Kong (e-mail: yufan.zhuang@connect.ust.hk).
			
			Shuai Wang is with the Shenzhen Institute of Advanced Technology, Chinese Academy of Sciences, Shenzhen 518055, China (e-mail: s.wang@siat.ac.cn).
			
			B.~Lv and R. Wang are with Department of Electrical and Electronic Engineering, Southern University of Science and Technology, Shenzhen 518055, China (e-mail: lyubj@mail.sustech.edu.cn, wang.r@sustech.edu.cn).

		 P. Xiao are with 5GIC \& 6GIC, Institute for Communication Systems (ICS), University of Surrey, Guildford GU2 7XH, UK (e-mail: p. xiao@surrey.ac.uk).

			%N. Li is with Beijing University of Posts and Telecommunications, Beijing, China (xiaonanahd163@163.com).  
			
		}%}
	
}

\maketitle

%\vspace{-0.5in}	

\begin{abstract}
	Most of the existing research on degrees-of-freedom (DoF) with imperfect channel state information at the transmitter (CSIT) assume the messages are private, which may not reflect reality as the two receivers can request the same content. To overcome this limitation, we therefore consider the hybrid unicast and multicast messages. In particular, we characterize the optimal DoF region for the two-user multiple-input multiple-output (MIMO) broadcast channel (BC) with imperfect CSIT and hybrid messages. For the converse, we establish a three-step procedure  to exploit the utmost possible relaxation. For the achievability, since the DoF region is with specific three-dimensional structure regarding antenna configurations and CSIT qualities, we verify the existence or non-existence of corner point candidates via the feature of antenna configurations and CSIT qualities categorization, and provide a hybrid message-aware rate-splitting scheme. Besides, we show that to achieve the strictly positive corner points, it is unnecessary to split the unicast messages into private and common parts. This  implies adding a multicast message may mitigate the rate-splitting complexity.
\end{abstract}


%reveal the shape of
%\bibliographystyle{IEEEtran}
%\bibliography{SDoFBC} 

\section{Introduction}
The emergence of the upcoming sixth generation mobile communications (6G) will provide extremely reliable, ultra-fast, and ubiquitous wireless connectivity with significantly
elevated performance, as opposed to
those in existing communication standards and systems \cite{Liyanaarachchi,Xinran,Xiaodan}. It is expected that  6G can achieve 50 times
higher peak data rate, 10 times reduced latency, and 100
times higher reliability than that of existing mobile communications systems. Typical 6G services are upgraded version of enhanced mobile broadband (eMBB), ultra reliable low latency communications (URLLC), and massive machine type communications (mMTC). One of the main challenges of 6G is the technique of multiple-access, which should support massive receivers with high data rate and are resilient to errors of channel state information at the transmitter (CSIT). Conventional multiple-access techniques, e.g., orthogonal multiple-access (OMA) and non-orthogonal multiple-access (NOMA), cannot address this challenge due to the following reasons: 1) OMA is very resource-consuming for supporting massive receivers 2) Both OMA and NOMA cannot be adaptive and robust to errors of CSIT, leading to degraded performance. In this regard, the rate-splitting multiple-access (RSMA) stands out as a viable solution, which not only   can  support massive receivers but also  is adaptive and robust against errors of CSIT \cite{3,A2,A3}. Specifically, it was found in \cite{3} that NOMA is in fact a special case of RSMA. The foundations of RSMA stem from information-theoretic research, and the concept of RSMA then further moves to engineering practice with advanced wireless communication applications. Next, we will review the  related work to the problem.

\subsection{Related Work}
The historical trajectory of RSMA from information-theoretic studies can be found in \cite{Carleial,han1981new,Etkin,Rimoldi,Medard,Sheng,  Tiangao, Xinping, PiovanoCL, linearly, Kaiming,Ye,Zhang,hao2017achievable,davoodi2020degrees,Anup,Hamdi,Piovano,HChenxi}. Date back as early as 1978, the idea of rate-splitting was proposed for the capacity of Gaussian interference channel (IC) \cite{Carleial}. 
In 1981, an advanced rate-splitting strategy with power and time sharing, known as Han-Kobayashi scheme, was initially proposed in Gaussian IC to establish a class of new capacity region \cite{han1981new}, where the degree-of-freedom (DoF) is increased by decoding part or whole of the interference.
In \cite{Etkin}, the rate-splitting-based strategy was shown to achieve the capacity region of
the deterministic interference channel within one bit. It was shown in \cite{Rimoldi} that any point in the capacity region of a Gaussian multiple-access channel is achievable by rate-splitting, i.e., each receiver ``splits" data and signal into two parts. Later, the authors of \cite{Medard} showed that by rate-splitting, the capacity region of slotted ALOHA multi-access systems is the same as the capacity region of multi-access with continuous  transmission.
% Considering the fast time-varying channel, the transmitter can only have perfectly delayed and imperfectly current CSIT. 

Despite this challenge, rate-splitting strategies  were shown to be DoF-optimal or constant-to-capacity \cite{Sheng,  Tiangao, Xinping, PiovanoCL, linearly, Kaiming,Ye,Zhang,hao2017achievable,davoodi2020degrees,Anup,Hamdi,Piovano,HChenxi}.  In particular, a rate-splitting scheme with delayed CSIT was first proposed in \cite{Sheng} for the two-user multi-input single-output (MISO) broadcast channel (BC). Then, an improved scheme with DoF-optimality was designed in \cite{Tiangao}. For the two-user MIMO BC,  the DoF region was characterized by rate-splitting in \cite{Xinping}. For the $K$-user MISO BC, the DoF region was given by \cite{PiovanoCL} via rate-splitting. It was shown in \cite{linearly} that the rate-splitting can achieve the capacity region within a constant gap in the two-user case. The authors of \cite{Kaiming} showed that rate-splitting achieves a constant gap to capacity region for a scalar Gaussian full-duplex cellular network with device-to-device (D2D) messages.
The rate-splitting design for information-theoretic security was first proposed in \cite{Ye} for the $K$-receiver MISO BC with imperfect CSIT. Then, the authors in \cite{Zhang} extended the idea in \cite{Ye} to the two-user MIMO BC with imperfect CSIT. In the case when messages are not confidential, the achievable DoF regions were derived via rate-splitting for the two receiver MIMO BC with arbitrary antenna configurations \cite{hao2017achievable}. Recently, it was found in \cite{davoodi2020degrees} that the achievable DoF region (DoF inner region) matches  the  DoF outer region, showcasing the superiority of rate-splitting. As a milestone, reference \cite{Anup} considered a general $K$-user symmetric MIMO BC scenario and derived the sum-DoF of RSMA, spatial division multiple access (SDMA) and MIMO NOMA for analysis and comparison; for perfect and imperfect CSIT, and different network loads. In \cite{Hamdi}, it was shown that rate-splitting achieves a higher symmetry DoF than SDMA for the $K$-user underloaded MISO BC with imperfect CSIT. Furthermore, it was shown in \cite{Piovano} that rate-splitting attains the complete DoF outer region of the $K$-user overloaded MISO BC with heterogeneous CSIT qualities. In \cite{HChenxi}, a rate-splitting scheme was shown to achieve the best known DoF region of the $K$-cell MISO IC with imperfect CSIT. 

The applications of RSMA in wireless systems can be found in \cite{4,5,6,7,8, VLC, Sandeep, Hassan2021clustering, Miaowen,  Mishra2022cellfree,   Chen2021multicast, Abolpour2022secrecy,   Shiyao}.  In \cite{4}, the integrated sensing and communication system was incorporated with rate-splitting. The authors in \cite{5} studied a joint design of intelligent reflecting surface (IRS) and rate-splitting, where the proposed framework outperformed the decode-and-forward rate-splitting. In \cite{6}, the authors showed that in contrast
to conventional multi-receiver and massive MIMO systems, for which performance
collapses under mobility, rate-splitting can maintain reliable multi-receiver
connectivity with mobility.  The authors in \cite{7} studied the rate-splitting in satellite systems, where they revealed the superiority of rate-splitting-based multigroup multicast beamforming in both terrestrial and multibeam satellite systems. In addition, the rate-splitting was investigated  in \cite{8}, with the aim of mitigating the inevitable hardware impairments in realistic massive MISO BC. In \cite{VLC},  the energy efficiency optimizations for both single-cell and multi-cell RSMA-based visible light communications were investigated. In \cite{Sandeep}, the integration of RSMA with unmanned aerial vehicle base station was investigated. In \cite{Hassan2021clustering}, a rate-splitting framework for multi-hop D2D was proposed which enables two D2D links to share certain orthogonal radio resource blocks by forming device-clusters. The authors in \cite{Miaowen} exploited the RSMA in a two-tier wireless backhaul heterogeneous network, where an efficient decentralized algorithm was developed.
In \cite{Mishra2022cellfree}, an RSMA-assisted downlink transmission framework for cell-free massive MIMO was proposed to ameliorate the effect of pilot contamination in the downlink and achieve a performance gain over a conventional cell-free massive MIMO network. The authors of \cite{Chen2021multicast} designed rate-splitting  precoders for an overloaded multicarrier multigroup multicast downlink system.
In the presence of untrusted receivers, the performance of rate-splitting considering outage probability and secrecy outage probability was analyzed in \cite{Abolpour2022secrecy}.   In \cite{Shiyao}, RSMA was applied to federated edge learning for latency minimization. 

However, all the above works only considered the private transmissions and overlooked the possibility that two receivers can request the same content, i.e., the impact of multicast message. The first work considering RSMA with non-orthogonal unicast and multicast messages, i,e., hybrid messages, transmission is \cite{Yijie-1}, where the authors proposed two
NOMA-assisted transmission strategies, and the precoders of all the strategies are optimized for the weighted sum rate or energy efficiency maximization and subject to the sum	power constraint and the quality of service rate requirements. 
Recently, in \cite{cuiying} and \cite{Yalcin}, the rate optimization problem for rate-splitting with hybrid messages was further studied under diverse application scenarios and user cases. For one application example, when simultaneous wireless streaming of a popular immersive video, in addition to unicast messages, there can be a multicast message desired by more-than-one users \cite{cuiying}. For another application example, 
in multi-group multi-beam satellite systems, a multicast message (not splitting from unicast messages) can be available for different terrestrial users \cite{Yalcin}. To date, the DoF region with imperfect CSIT has not yet been investigated with hybrid messages, even for the two-user MIMO BC.

\subsection{Contributions}
In this paper, to overcome the above limitation, we consider the hybrid messages for the two-user $(M,N_1,N_2)$ MIMO BC with imperfect CSIT, where the transmitter has $M$ antennas, the receiver $k, k=1,2$ has $N_k$ antennas.   Tight converse and achievability proofs for the DoF region are given. Compared with \cite{hao2017achievable}, the impact of multicast message on rate-splitting of unicast messages is considered.  Our main contributions  are summarized as follows.

\begin{itemize}
	\item \textit{Corner Points}. We reveal the expression of all corner points for  the DoF region of the two-user MIMO BC with hybrid messages and imperfect CSIT. Since the DoF region has a specific three-dimensional structure w.r.t. antenna configurations and CSIT qualities,  existence verification of corner points, especially for the strictly positive corner points, is a challenging task. By dividing the antenna configurations and CSIT qualities into cases, we therefore verify the existence or non-existence of corner point candidates via the feature of antenna configurations and CSIT qualities.

	\item \textit{Achievability}.  We derive the achievability for the DoF region of the two-user MIMO BC with hybrid messages and imperfect CSIT. This achievability proof is given by showing corner points are achievable. To this end,  utilizing the space-time rate-splitting transmission scheme provided by \cite{hao2017achievable}, we provide a hybrid message-aware rate-splitting scheme, which relies on the power allocation and cases division in \cite{hao2017achievable}.  Furthermore, we show that to achieve strictly positive corner points, splitting the unicast messages into private and common parts is unnecessary, and the allocated power for the common part should be zero. This implies that adding a multicast message may mitigate the rate-splitting complexity of unicast messages.
	
	\item \textit{Converse}. We establish the  converse for DoF region of the two-user MIMO BC with hybrid messages and imperfect CSIT. In particular, we prove the  converse by relaxing the decodability of receivers and thus enhancing the original channel. The converse proof has three steps, i.e., relaxation of receiver one, relaxation of receiver two, and union of them. We then show  that this converse indeed exploits the utmost of possible relaxation by the matching with the proposed achievability.
	
	%\item We obtain the DoF region of the two-user MIMO BC with hybrid messages and imperfect CSIT. Unlike the DoF region with unicast message}s, we show that this DoF region is with specific three-dimensional structure w.r.t. different antenna configurations.  The interactions of  private and multicast message}s are characterized by the DoF region and related corner points. We find that to achieve the strictly positive corner points of the DoF region, splitting the unicast message}s into private and common parts is unnecessary, because the allocated power for the common part of unicast message}s should be zero. This implies that adding a multicast message} may mitigate the rate-splitting burden of unicast message}s, thereby  the overall implementation complexity. Moreover, it can be seen that the sum-DoF with hybrid messages is equal to that with unicast message}s only, since adding multicast message} does not create extra spatial and signal spaces.  Additionally, in the presence of delayed imperfect CSIT, we characterize the DoF region of the two-user MIMO BC with  hybrid messages.
\end{itemize}

%The remainder of this paper is organized as follows. 
\subsection{Organization \& Notation}
We introduce our system model in Section-II. Then, we summarize and discuss our main results  in Section-III. The achievability of Theorem 1 is given in Section-IV. The converse  of Theorem 1  is provided in Section-V. The proof of Theorem 2 is given in Section-VI. We discuss the extension to $K$-user setups in Section-VII. Finally, we draw our conclusions in 
Section-VIII.

The notation of this paper is given as follows. $a$, $\textbf{a}$, $\textbf{A}$ denote a scalar, a vector, and a matrix, respectively.  $\mathbb{E}\{\cdot\}$ denotes the long-term expectation operator.  The identity matrix with $M$ dimensions is denoted by $\textbf{I}_{M}$. Furthermore, the definitions of operations and specific symbols are summarized in Table I. 

\begin{table}[t]
\centering
\caption{Definitions of Notation}
\begin{tabular}{l|l}
\hline
\textbf{Symbol} & \textbf{Definition}  \\ \hline
$M$ & Number of antenna at $\text{Tx}$ \\ \hline
$N_k,k=1,2$ & Number of antenna at $\text{Rx}_k$  \\ \hline 
$P$ & Transmit power  \\ \hline 
$f(\cdot)$ & Encoding function  \\ \hline 
$g(\cdot)$ & Decoding function  \\ \hline 
$W_{pk},k=1,2$ & \makecell[l]{Private part of unicast messages for   $\text{Rx}_k$} \\ \hline 
$W_{c}$ & \makecell[l]{Common message merged} \\ \hline
$W_0$ & Multicast message to be transmitted \\ \hline
$\textbf{H}_k,k=1,2$ & \makecell[l]{The channel from the transmitter to $\text{Rx}_k$}\\ \hline 
$\hat{\textbf{H}}_k,k=1,2$ & The imperfect CSIT between Tx and $\text{Rx}_k$ \\ \hline
$\alpha_k,k=1,2$ & The CSIT quality \\ \hline
$\mathbf{n}_k,k=1,2$ & AWGN signal at  $\text{Rx}_k$ \\ \hline
$\textbf{s}$ & Transmit signal \\ \hline 
$\textbf{y}_k,k=1,2$ & Received signal at  $\text{Rx}_k$   \\ \hline
$d_k,k=1,2$ & DoF of unicast messages intended for  $\text{Rx}_k$ \\ \hline  
$d_0$ & DoF of multicast message  \\ \hline  
$\textbf{u}_k,k=1,2$ &  Private symbols for  $\text{Rx}_k$  \\ \hline  
$\textbf{u}_0$ &  Split common symbols for two receivers \\ \hline  
$\textbf{c}$ & Multicast symbols for two receivers \\ \hline 
$(\cdot)^H$ & Hermitian of a matrix ot vector\\ \hline 
$(\cdot)^T$ & Transpose of a matrix or vector \\ \hline
$(\cdot)^{\perp}$ & Null space of a matrix or vector\\ \hline
$[a]^+$ & $\text{max}\{a,0\}$\\ \hline
\end{tabular}	
\end{table}

%\begin{figure}[t]
%\centering
%	\includegraphics[width=2.5in]{Org}
%	\caption{Illustration of organization of this paper.}
%\end{figure}scenario
\begin{figure}[t]
\centering
\includegraphics[width=0.325\textwidth]{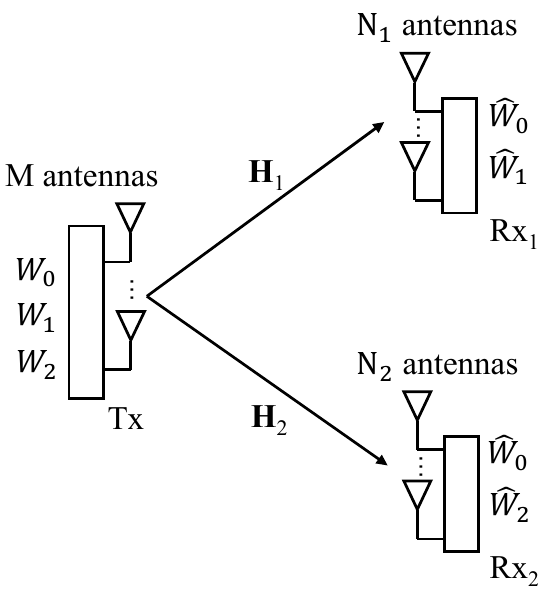}
\caption{The two-user $(M,N_1,N_2)$ MIMO BC with hybrid messages.}
\end{figure}
\section{System Model}

%    \begin{figure}[t]
%    	\centering
%    	\includegraphics[width=2in]{12.9}
%    	\caption{Illustration of the scenario of the two-user $(M,N_1,N_2)$ MIMO BC with hybrid messages.}
%    \end{figure}

We consider a two-user $(M,N_1,N_2)$ MIMO BC with an $M$-antenna transmitter, denoted by $\text{Tx}$, a receiver with $N_1$ antennas denoted by $\text{Rx}_1$ and a receiver with $N_2$ antennas denoted by $\text{Rx}_{2}$, which is illustrated in Fig. 1. The transmitter has unicast messages $W_1$ and $W_2$ for $\text{Rx}_1$ and $\text{Rx}_2$, respectively, and a multicast message $W_0$ for both two receivers. Mathematically, the signal received at $\text{Rx}_k, k=1,2$ can be written as
\begin{equation}
\mathbf{y}_k=\mathbf{H}_k^H\mathbf{s}+\mathbf{n}_k,
\end{equation}
where $\mathbf{s}$ denotes the transmitted symbol, $\mathbf{n}_k \sim \mathcal{CN}(0,\mathbf{I}_{N_k})$ denotes the AWGN vector at $\text{Rx}_k$, $\mathbf{H}_k \in \mathbb{C}^{M \times N_k}$ denotes the channel matrix between Tx and $\text{Rx}_k$. $\mathcal{H} := \{\mathbf{H}_1,\mathbf{H}_2\}$  follows an ergodic stationary process with probability density $f_{H}(\mathcal{H})$, and assumed
to be stationary and ergodic.

If no other statements, we consider the case with perfect channel state information at receivers (CSIR) while imperfect CSIT \cite{JoudehClerckx}. Specifically, at the receiver side, the channel gains can be easily obtained by sending a pilot sequence for channel estimation. That is, Rx$_k$ knows the precoders and channel matrices to decode the desired signal. At $\text{Tx}$, there exists imperfect CSIT resulting from channel estimation errors, quantization errors, prediction errors, etc.  Let $\widehat{\textbf{H}}_k$ denote the impefect CSIT, where $\textbf{H}_k = \widehat{\textbf{H}}_k + \widetilde{\textbf{H}}_k$ and $\widetilde{\textbf{H}}_k$ denotes the estimation error matrix. Furthermore, $\alpha_{k} \ge 0$ denotes CSIT quality. Henceforth, we focus $0\leq\alpha_{k}\leq1$, because $\alpha_{k}\geq1$ is equivalent to perfect CSIT, where the interference will be dissolved in noise, and $\alpha_{k}=0$ implies no CSIT, where the interference has the same power level as  desired signal \cite{hao2017achievable}.

%{\color{blue}One approach to exploit the capacity of MIMO system is to employ spatial multiplexing. The imformation streams can be seperated by precoding techniques such as maximum likelihood (ML) which achieves optimal performance or linear receivers like Zero-Forcing (ZF) which provide sub-optimal performance but offer significant computational complexity reduction with tolerable performance degradation. However these techniques require accurate channel state informaiton.\cite{wang2007performance}}

Furthermore, we focus on the case with hybrid message, i.e., $(W_1,W_2,W_0)$. The multicast message intended for both receivers is denoted by $W_0$.  Generally, the encoding function at $\text{Tx}$ is expressed as
%\begin{equation}
%    	\mathbf{s}=f(W_0,W_1,W_2,W, \widehat{\textbf{H}}_1,\widehat{\textbf{H}}_2),
$\mathbf{s}=f(W_0,W_1,W_2,\widehat{\textbf{H}}_1,\widehat{\textbf{H}}_2)$,
%\end{equation} 
where $f$ is designed specifically in Section-IV for our problem. Note that we will discuss later on the delayed CSIT definition, which may exist for the time-varying channel.
%, assuming that the channel is time-varying, it is defined as the CSIT only contains channel state information (CSI) in past time slots but not the CSI in current time slot.

The decoding function at $\text{Rx}_k$, denoted by $g(\cdot)$, decodes $(\widehat{W}_k,\widehat{W}_0) = g(\textbf{y}_k,\textbf{H}_1,\textbf{H}_2)$. Let $R_0$ denote the rate of multicast message, and let $R_k, k=1,2,$ denote the entire rate of unicast message. The rate is said to be achievable, if there are a sequence of codebook pairs $\{\mathcal{B}_{1,t},\mathcal{B}_{1,t}\}_{t=1}^n$ and decoding functions $\{g_{1,n},g_{2,n}\}$ such that the error probabilities $\mathcal{P}_{e}^{[n]}(\widehat{W}_i \ne W_i),\forall i$ go to zero when $n$ goes to infinity. The capacity region, denoted by $\mathcal{C}(\rho)$, where $\rho$ is defined as signal-noise-ratio (SNR), is the region of all such achievable rate tuples. According to \cite{hao2017achievable} and \cite{davoodi2020degrees}, the achievable DoF is defined as $d = \lim_{\rho \rightarrow {\cal{1}}} \frac{R(\rho)}{\log \rho}$. The DoF region is defined as the pre-log factor of the capacity region as $\rho \rightarrow {\cal{1}}$,
\begin{equation} 
\mathcal{D} :=	\left\{ 
(d_{1}, d_{2}, d_0) \in \mathbb{R}_+^3 \left| 
\begin{split} 
(R_{1}(\rho), R_{2}(\rho),R_{0}(\rho)) \in \mathcal{C}(\rho), \\
d_{i} = \lim_{\rho \rightarrow {\cal{1}}} \frac{R_{i}(\rho)}{\log \rho}, i = 0,1,2.
\end{split}\right.\right\}. \nonumber 
\end{equation}
where $d_0, d_1, d_2$ denote the DoF of multicast message, and unicast messages for $\text{Rx}_1$ and $\text{Rx}_2$, respectively.

\section{Main Results and Discussion}

%In this section, we summarize the mains results of this paper, including the DoF region and sum-DoF of the  two-user $(M,N_1,N_2)$ MIMO BC with hybrid messages and imperfect CSIT. We also discuss the implications of the main results. 

\begin{theorem}[DoF Region]
For the two-user $(M,N_1,N_2)$ MIMO BC with hybrid messages and imperfect CSIT, the DoF region, denoted by $\mathcal{D}$, is given below. 
\begin{eqnarray}\label{region} 	
\mathcal{D} = 		\left\{	(d_1,  d_2,   
d_0)\in \mathbb{R}_+^3 
|  \right.  \qquad  \qquad \qquad \qquad \qquad \qquad \,\,\,\,\, \nonumber  \\	  		
\left. 	\begin{aligned} 		 		 
&	\,\,\,\,\,	d_1+d_0 \le \min\{M,N_1\}, \\
&	\,\,\,\,\,  d_2+d_0 \le \min\{M,N_2\}, \\
&	\,\,\,\,\,	d_1+d_2+d_0 \le \min\{M,N_2\} + \\
& \qquad \quad \,\,  [\min\{M,N_1+N_2\} - \min\{M,N_2\}] \alpha_0,  \\
&	\,\,\,\,\,	\frac{d_1+d_0}{\min\{M,N_1\}} + \frac{d_2}{\min\{M,N_2\}} \leq
1 + \\
&  \qquad \quad \,\, \frac{\min\{M,N_1+N_2\} - \min\{M,N_1\}} {\min\{M,N_2\}} \alpha_1.   	
\end{aligned} \right\}, 
%\right\}, 
\end{eqnarray} 	
where $\alpha_0$ is given in \cite[eqn. (8)]{hao2017achievable} and copied on the top of this page.

\begin{figure*}

\begin{equation}
\alpha_0 = \begin{cases}
\alpha_2, & \text{if } \Phi \le 0, \\
\alpha_2 - \dfrac{\Phi}{\min\{M,N_1+N_2\} - \min\{M,N_1\}}, & \text{else if } \alpha_1 \ge 1 - \alpha_2, \\
\dfrac{\alpha_1 \alpha_2\left[\min \left\{M, N_1+N_2\right\}-\min \left\{M, N_2\right\}\right]}{\left[\min \left\{M, N_2\right\}-\min \left\{M, N_1\right\}\right]\left(1-\alpha_1\right)+\left[\min \left\{M, N_1+N_2\right\}-\min \left\{M, N_2\right\}\right] \alpha_2}, & \text{else if } \alpha_1 \le 1 - \alpha_2.
\end{cases}
\end{equation}
where
\begin{equation}
\Phi := \min \left\{M, N_2\right\}-\min \left\{M, N_1\right\}+\left[\min \left\{M, N_1+N_2\right\}-\min \left\{M, N_2\right\}\right] \alpha_2-\left[\min \left\{M, N_1+N_2\right\}-\min \left\{M, N_1\right\}\right] \alpha_1.
\end{equation}
\hrule

\end{figure*}

\end{theorem} 
\begin{IEEEproof}
The achievability proof is presented in Section-IV. Subsequently, the converse proof is given in Section-V.  
\end{IEEEproof}

\begin{remark}[What Do Hybrid Messages Change?] This work, undoubtedly, the main differences from the DoF region of the two-user MIMO BC with unicast messages \cite{hao2017achievable} and the DoF region of the $K$-user MISO BC with unicast messages \cite{PiovanoCL} are DoF region and corner points, where the proposed DoF region is 3-dimensional and a large parts of corner points are new. In particular, the proposed DoF region characterizes the interplay of unicast messages, multicast message, antenna configurations, and CSIT qualities, under limited signal space and shared spatial domain.  The emerging strictly positive corner points of the  DoF region capture the hybrid unicast and multicast messages under imperfect CSIT. Several strictly positive corner points  exist only if some conditions hold. Furthermore, it can be seen from our proposed scheme that to achieve the strictly positive corner point of the DoF region, there is no need in splitting the unicast messages into private and common parts, as this point we will see later on. 
\end{remark}

\begin{figure}[t]
\centering
\includegraphics[width=0.5\textwidth]{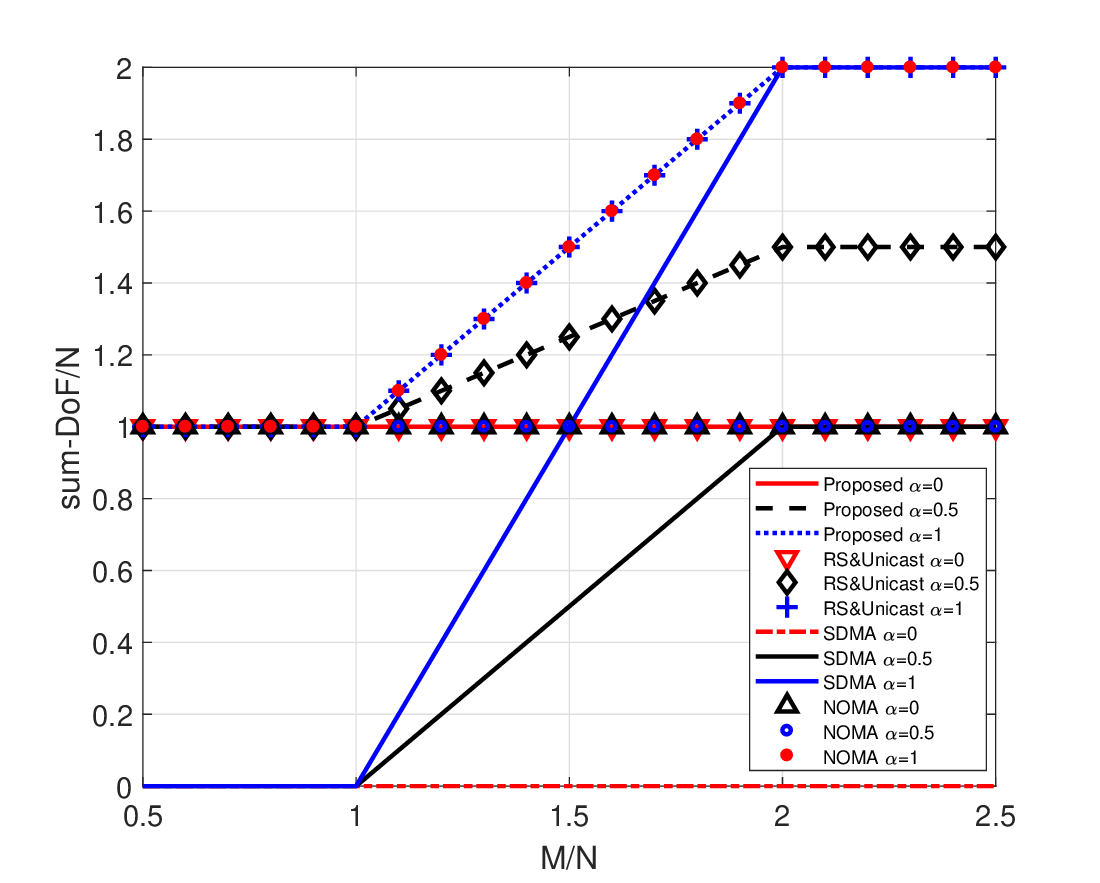}
\caption{Proposed v.s. competing schemes with or without rate-splitting.}
\label{F2} 
\end{figure}

\begin{remark}[Sum-DoF]
According to Theorem 1, the sum-DoF of this $(M,N_1,N_2)$ MIMO system, defined in Section-II, is given as follows.	
\begin{enumerate}
\item $\sum_{i=0}^2 d_i = N_2+(M-N_2)\alpha_{2}$ on the condition that $\frac{N_2-N_1+(M-N_2)\alpha_{2}}{M-N_1} \leq \alpha_{1}$.
\item $\sum_{i=0}^2 d_i = \max\{N_1 + (M-N_1)\alpha_1$, $ N_2+(M-N_2)(-\frac{N_2-N_1}{M-N_1}+\frac{N_2-N_1}{M-N_1}\alpha_{2}+\alpha_{1})\}$ on the condition that $1 - \alpha_2 \leq \alpha_{1} <  \frac{N_2-N_1+(M-N_2)\alpha_{2}}{M-N_1}$. 
\item $\sum_{i=0}^2 d_i = \max\{	\frac{(M-N_2)^2\alpha_{1}\alpha_{2}}{(M-N_2)\alpha_{2}+(N_2-N_1)(1-\alpha_{1})}+N_2, (M-N_1)\alpha_1 + N_1\}$ on the condition that $\alpha_{1} \leq 1 - \alpha_2$.
\end{enumerate}  		  
The sum-DoF with $N_1=N_2=N$ and different CSIT qualities is illustrated in Fig. \ref{F2}. We compared the proposed scheme with \textbf{RS}\&\textbf{Unicast scheme} (i.e., rate-splitting with unicast messages \cite{hao2017achievable}), \textbf{NOMA scheme} (i.e., NOMA  without rate-splitting  \cite{OJCOM}), and \textbf{SDMA scheme} (i.e., spatial zero-forcing (ZF) transmission without rate-splitting \cite{OJCOM}). It can be seen that the proposed sum-DoF is the same as that of \textbf{RS}\&\textbf{Unicast scheme}. This is because, although the dimension elevation of the DoF region when a multicast message is additionally considered, supplying a multicast message does not create extra spatial or signal space. It can be further seen that the proposed scheme achieves a higher sum-DoF than \textbf{NOMA scheme}, confirming the results in \cite{OJCOM}. Moreover, it shows that except perfect CSIT (i.e., $\alpha = 1$) and $M \ge 2N$, due to the gain brought by the common message, the proposed scheme achieves a higher sum-DoF than \textbf{SDMA scheme}.
%	\begin{figure}[t] 
%			\centering
%			\includegraphics[width=8cm]{sumDoF.eps}
%			\caption{Sum-DoF with hybrid messages and different CSIT qualities.}\label{F2}
%			\end{figure}
\end{remark}		

%	\begin{figure}[t] 
%	\centering
%	\includegraphics[width=8cm]{Delayed}
%	\caption{The DoF region with delayed and imperfect CSIT, and private and multicast message}s, when $M=2$, $N_1=N_2=1$, and $\alpha_1=\alpha_2=0.5$.} \label{DE}
%\end{figure}	

The delayed CSIT does not reflect the current CSI but does match with the past CSI. Mathematically, delayed CSIT with errors can be defined as follows. The imperfect CSIT matrices evolves randomly with time, which can be modeled as $\widehat{\textbf{H}}_k[t]$ for time slot $t$. Under the assumption of delayed CSIT with errors, the encoding function at $\text{Tx}$ and time slot $t$ is given by $f_t(W_0,W_1,W_2,\widehat{\textbf{H}}_1[1],\cdots,\widehat{\textbf{H}}_1[t-1],\widehat{\textbf{H}}_2[1],\cdots,\widehat{\textbf{H}}_2[t-1])$. The difference from encoding function with imperfect CSIT in Section-II is that we assume the channel is time-varying and can only utilize past CSI matrices with errors. 

\begin{theorem}[Delayed CSIT with Errors]
For the two-user $(M,N_1,N_2)$ MIMO BC in the presence of delayed CSIT with errors, and hybrid messages, the DoF region, denoted by $\mathcal{Q}$, is given below.
\begin{eqnarray}
\mathcal{Q} = \{(d_1,d_2,d_0)  
\in \mathbb{R}_+^3 | \qquad \qquad \qquad \qquad \qquad \qquad \nonumber \\
\left.
\begin{aligned}
\frac{d_1}{\min\{N_1+\alpha_2N_2,M\}} 
+ \frac{d_2+d_0}{\min\{N_2,M\}} \le 1, \\
\frac{d_1+d_0}{\min\{N_1,M\}} + \frac{d_2}{\min\{N_2+\alpha_1N_1,M\}}   \le 1.
\end{aligned}
\right\}. \label{Delayed}
\end{eqnarray}
\end{theorem}
\begin{IEEEproof}
The converse and achievability proofs are provided in Section VI.
\end{IEEEproof}	

\begin{remark}[Comparison]		
\begin{figure}[t]
\centering  %图片全局居中
%\subfigbottomskip=2pt %两行子图之间的行间距
%\subfigcapskip=-5pt %设置子图与子标题之间的距离
\subfigure[]{
\includegraphics[width=0.46\linewidth]{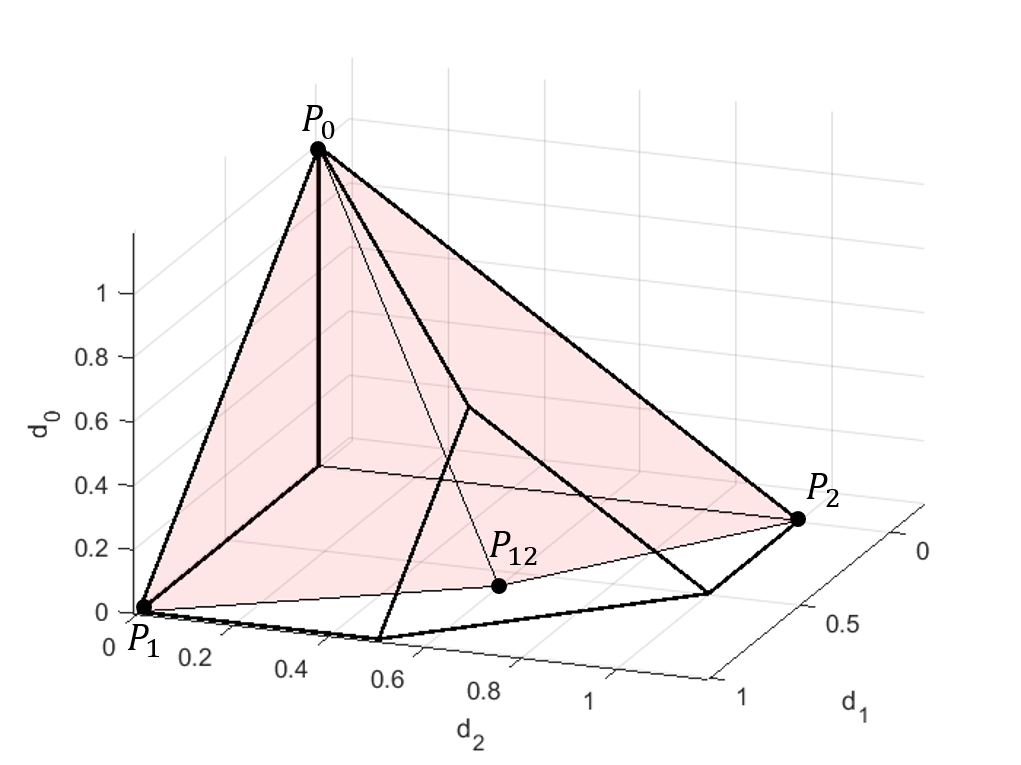} \label{FD1}} 
\subfigure[]{
\includegraphics[width=0.46\linewidth]{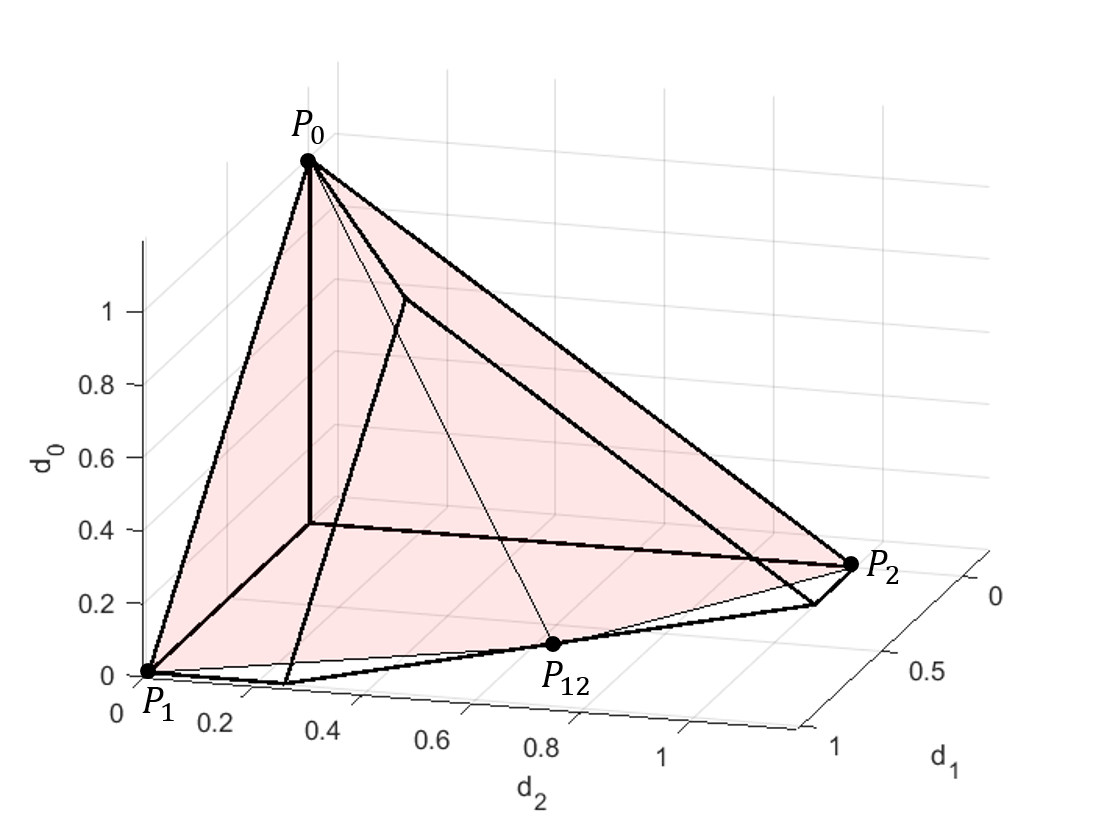}\label{FD2}} \\			 
\subfigure[]{
\includegraphics[width=0.46\linewidth]{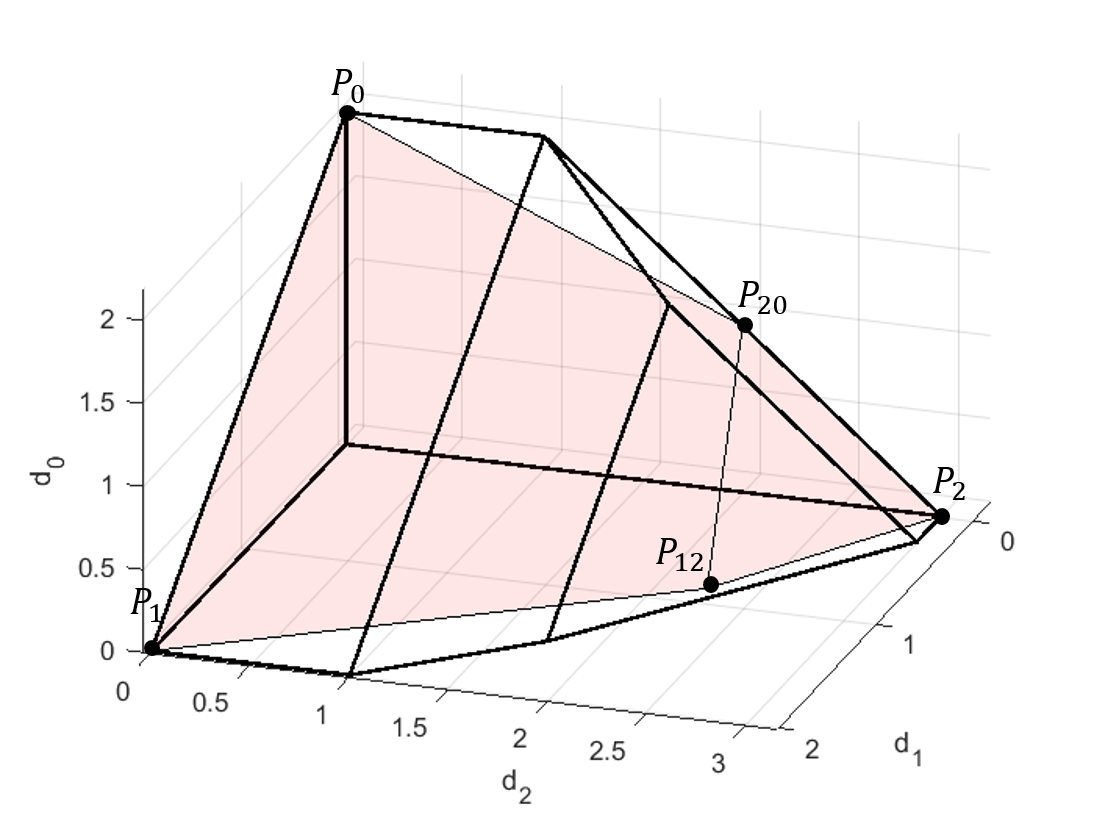}\label{FD3}}
%\quad
\subfigure[]{
\includegraphics[width=0.46\linewidth]{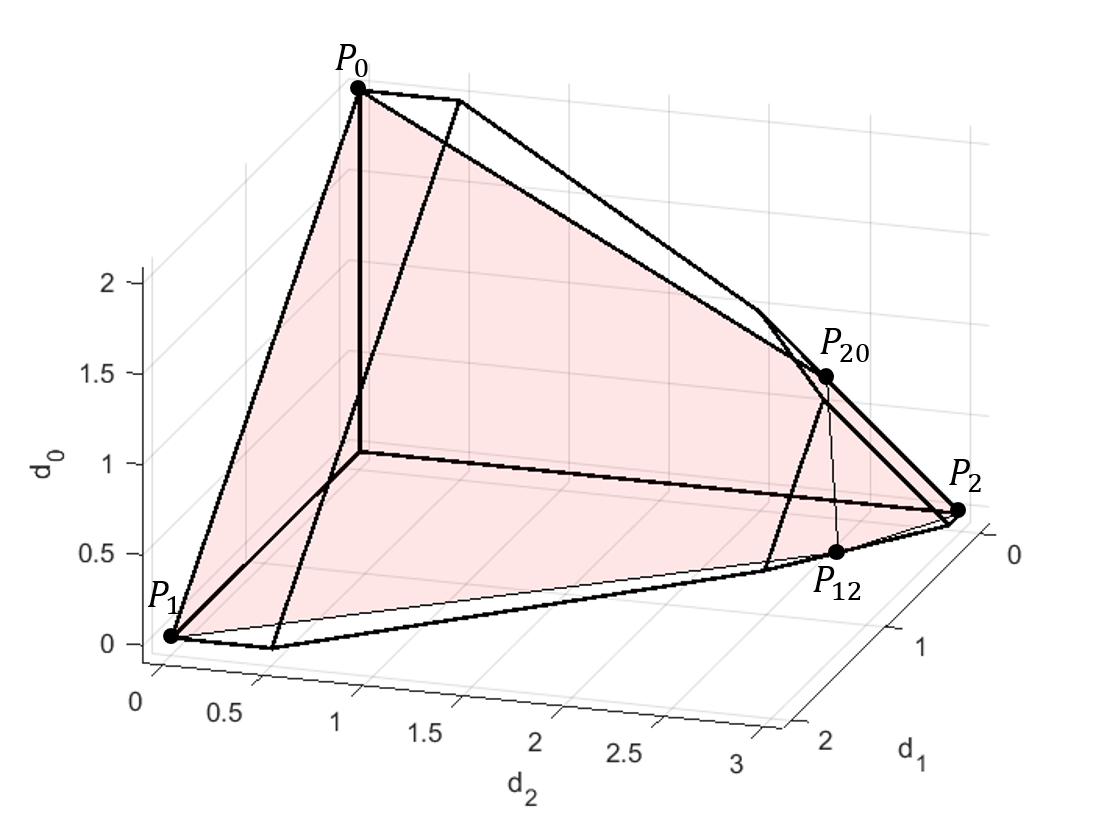}\label{FD4}} 
\caption{a) $M=2$, $N_1=N_2=1$,  $\alpha_1=\alpha_2=0.5$, b) $M=2$, $N_1=N_2=1$,  $\alpha_1=0.25, \alpha_2=0.75$, c) $M=4$, $N_1=2$, $N_2=3$, $\alpha_1=\alpha_2=0.5$, and d) $M=4$, $N_1=2$, $N_2=3$, $\alpha_1=0.25, \alpha_2=0.75$. The hull of the DoF region with delayed CSIT in the presence of errors is blackened.}	
\end{figure}
It worth mentioning that delayed CSIT with errors may occur when the CSI feedback is lagging behind the variations of the channel, and the feedback has distortion. To show the difference, we compare the DoF region of imperfect CSIT \eqref{region} with the DoF region of delayed CSIT with errors  \eqref{Delayed} in Fig. 3 with 4 parameters. It can be seen from Fig. 3 that the DoF region of delayed CSIT with errors is contained in the DoF region of imperfect CSIT, for the considered parameters. This shows the hindering effect of delayed CSIT, compared with no delayed one.

\end{remark}

\section{Achievabiliy Proof of Theorem 1: Proposed System and Scheme Design}

%To show the achievability, we begin with an illustrative example and then present the unified framework.

In this section, we first present the hybrid message-aware rate-splitting system, and then a unified hybrid message-aware rate-splitting scheme. Finally, we provide an example to help readers better grasp the design principle and insights.

\subsection{Proposed Hybrid Message-Aware Rate-Splitting System}
%In this subsection, in order to achieve the DoF region, we therefore propose a hybrid message-aware rate-splitting system as a precondition.

This system adopts ZF and rate-splitting  to construct the transmission procedure. In particular, $\mathbf{w}_k \in \mathbb{C}^{M\time1}$ denotes the ZF precoder that is a unit norm vector in the null space of $\widehat{\textbf{H}}_k$. Then, if the transmitted signal is ZF-precoded, the strength of the residual interference at the unintended receiver can be written as  $|\mathbf{h}_{k,i}^H\mathbf{w}_k|^2$, where $\mathbf{h}_{k,i}$ denotes the $i^\text{th}$ column of $\mathbf{H}_{k}$.  Note that $\mathbb{E}\{|\mathbf{h}_{k,i}^H\mathbf{w}_k|^2\} \sim P^{-\alpha_k}$.
Also, the rate-splitting technique is adopted.   Specifically, the message $W_k$ intended for Rx$_k$ is split into a private part $W_{pk}$ decoded by Rx$_k,$ $k=1,2$ and a common part $W_{ck}$ drawn from a shared codebook decoded by two receivers. Then, the common parts $W_{c1}$ and $W_{c2}$ from two receivers' unicast messages merges as a common message $W_c$. Finally, the common message $W_c$ is combined with multicast message $W_0$,  recasting as a composite multicast symbol \cite{Yijie-1}.  

The decoding procedure is given as follows. The composite multicast symbol is first decoded. Next, successive interference cancellation (SIC) is used to cancel the interference aroused by composite multicast symbol. Finally, the unicast symbol is decoded separately at each receiver.

\subsection{A Unified Hybrid Message-Aware Rate-Splitting Scheme}
In this subsection,  based on the proposed system and utilizing the space-time rate-splitting transmission scheme provided by \cite{hao2017achievable}, we are able to crystallize a hybrid message-aware rate-splitting scheme. Note that our scheme belongs to 1-layer rate-splitting, which is different from 2-layer rate-splitting in the species of common messages. Without loss of generality, we consider $N_1 \leq N_2 \leq M$. Furthermore, we assume $M\leq N_1+N_2$, since for other cases the DoF region can be achieved by turning off redundant transmit or receive antennas. This is because the DoF region of $M > N_1+N_2$ and the DoF region of $M\leq N_1+N_2$ are the same. Under $N_1 \leq N_2 \leq M\leq N_1+N_2$, the DoF region is simplified to 
\begin{eqnarray}
\mathcal{D} = \{(d_1,d_2,d_0)  
\in \mathbb{R}_+^3| \qquad \qquad \qquad \qquad \qquad \,\,\, \nonumber \\	
\left.	\begin{aligned}
&	\ell_1: d_1+d_0 \le N_1, \\
&	\ell_2: d_2+d_0 \le N_2, \\
&	\ell_3: d_1+d_2+d_0 \le N_2+(M-N_2)\alpha_0,  \\
&	\ell_4: \dfrac{d_1+d_0}{N_1} + \dfrac{d_2}{N_2} \leq 1+\dfrac{M-N_1}{N_2}\alpha_1.  
\end{aligned}
\right\}, 
\end{eqnarray}
where $\alpha_0$ is given in \cite[eqn. (8)]{hao2017achievable} and can be seen in (3).

Henceforth, we constitute the rate-splitting transmission block for this two-user $(M,N_1,N_2)$ MIMO BC with hybrid messages and imperfect CSIT as follows.
\begin{itemize}
\item $M-N_2$ unicast symbols, denoted by $\mathbf{u}_1 \in \mathbb{C}^{(M-N_2) \times 1}$, are transmitted to $\text{Rx}_1$ along a ZF-precoder $\mathbf{V}_1=\widehat{\textbf{H}}_2^{\perp} \in \mathbb{C}^{M \times (M-N_2)}$ with power exponent $A_1$;

\item $M-N_1$ unicast symbols, denoted by $\mathbf{u}_2^{1} \in \mathbb{C}^{(M-N_1) \times 1}$, are transmitted to $\text{Rx}_2$ along a ZF-precoder $\mathbf{V}_2^{1}=\widehat{\textbf{H}}_1^{\perp} \in \mathbb{C}^{M \times (M-N_1)}$ with power exponent $A_2$;

\item $[N_1+N_2-M]^+$ unicast symbols, denoted by $\mathbf{u}_2^{2}$, is transmitted to $\text{Rx}_2$ along a precoder $\mathbf{V}_2^{2} \in \mathbb{C}^{M \times [N_1+N_2-M]^+}$ in the subspace spanned by $\widehat{\textbf{H}}_2$ with power exponent $[A_2-\alpha_1]^+$;

\item The  composite multicast symbol, denoted by $(\mathbf{c}+\mathbf{u}_0) \in \mathbb{C}^{M \times 1}$, is multicast with remaining power, where $\mathbf{c}$ denotes the common parts split from the unicast messages and $\mathbf{u}_0$ denotes the multicast symbol to be transmitted.
\end{itemize}

Moreover, the power exponents $A_1$ and $A_2$ are defined as $A_1\in[0,\alpha_2]$ and $A_2\in[0,1]$. Mathematically, the transmitted and received signals are written as
\begin{subequations}
\begin{equation}
\mathbf{s}=\underbrace{\mathbf{c+u_0}}_{\sim P}+\underbrace{\mathbf{v}_{1} \mathbf{u}_{1}}_{\sim P^{A_{1}}}+\underbrace{\mathbf{V}_{2}^{1} \mathbf{u}_{2}^{1}}_{\sim P^{A_{2}}}+\underbrace{\mathbf{v}_{2}^{2} \mathbf{u}_{2}^{2}}_{\sim P^{\left[A_{2}-\alpha_{1}\right]^{+}}}, 
\end{equation}
\begin{equation}
\mathbf{y}_{1}=\underbrace{\mathbf{H}_{1}^{H} \mathbf{(c+u_0)}}_{\sim P} + \underbrace{\mathbf{H}_{1}^{H} \mathbf{v}_{1} \mathbf{u}_{1}}_{\sim P^{A_{1}}}+\underbrace{\mathbf{H}_{1}^{H}\left(\mathbf{V}_{2}^{1} \mathbf{u}_{2}^{1}+\mathbf{v}_{2}^{2} \mathbf{u}_{2}^{2}\right)}_{\sim P^{\left[A_{2}-a_{1}\right]^{+}}}, 
\end{equation}
\begin{equation}
\mathbf{y}_{2}=\underbrace{\mathbf{H}_{2}^{H} \mathbf{(c+u_0)}}_{\sim P} + \underbrace{\mathbf{H}_{2}^{H} \mathbf{v}_{1} \mathbf{u}_{1}}_{\sim P^{A_1-\alpha_{2}}}+\underbrace{\mathbf{H}_{2}^{H} \mathbf{V}_{2}^{1} \mathbf{u}_{2}^{1}}_{\sim P^{A_{2}}}+\underbrace{\mathbf{H}_{2}^{H} \mathbf{v}_{2}^{2} \mathbf{u}_{2}^{2}}_{\sim P^{\left[A_{2}-\alpha_{1}\right]^{+}}}.
\end{equation}
\end{subequations}

As we can see from the received signals, if $A_2 \leq \alpha_1$, the undesired unicast symbols are dissolved into the noise. If $A_2 \textgreater \alpha_1$, the designed power allocation policy will ensure that all the three unicast symbols intended for $\text{Rx}_2$ are received by $\text{Rx}_1$ with the same power level. Similar to \cite{hao2017achievable}, for (8b) and (8c), utilizing the proof in \cite[Appendix A]{hao2017achievable} and considering that each receiver decodes the common part splitting  from the unicast messages and the multicast message successively, the following DoF tuple is achievable.

At $\text{Rx}_1$, we have
\begin{subequations}
\begin{eqnarray}\label{UQ1}	 
&& \!\!\!\!\!\!\!\!\!\!\!\!	d_{0}+d_{c} \leq d_c^1 := N_1 - (M-N_2)\max\left\{A_{1}, A_{2}-\alpha_{1}\right\}   \nonumber \\
&& \!\!\!\!\!\!\!\!\!\!\!\! \qquad \qquad \qquad \quad \,\, - (N_1+N_2-M)\left[A_{2}-\alpha_{1}\right]^{+}, \\	 
\label{UQ2}
&& \!\!\!\!\!\!\!\!\!\!\!\! d_{p1} \le (M-N_2)\left[A_{1}-\left[A_{2}-\alpha_{1}\right]^{+}\right]^{+},
\end{eqnarray}
where $d_c$ and $d_{p1}$ denote achievable DoF for $W_c$ and $W_{p,1}$, respectively.

At $\text{Rx}_2$, we have
\begin{eqnarray}\label{UQ3} 
&&	\!\!\!\!\!\!\!\!\!\!\!\! d_{0}+d_{c}\leq d_c^2:= N_2 - (M-N_1)A_{2} \nonumber \\ 
&& \!\!\!\!\!\!\!\!\!\!\!\! \qquad \qquad \qquad \quad \,\, - (N_1+N_2-M) \left[A_{2}-\alpha_{1}\right]^{+},	
\\
&&	\!\!\!\!\!\!\!\!\!\!\!\! d_{p 2} \le (M-N_1)A_{2} + (N_1+N_2-M)\left[A_{2}-\alpha_{1}\right]^{+}, \label{UQ4}
\end{eqnarray}
\end{subequations}
where $d_{p2}$ denotes the DoF for $W_{p,2}$.

Accordingly, the achievable sum-DoF is defined as 
\begin{equation}\label{sumUDoF}
d_{s}(A_1,A_2):= \min\{d_s^{1}(A_2),d_s^{2}(A_1,A_2)\},
\end{equation}
where
\begin{subequations}
\begin{eqnarray}\label{sumUDoF1}
d_s^1\left(A_{2}\right)=N_1 + (M-N_1)A_{2} - (M-N_2)\left[A_{2}-\alpha_{1}\right]^{+}, \label{sumUDoF2} \nonumber \\
d_s^2\left(A_{1}, A_{2}\right) = N_2 + (M-N_2)\left[A_{1}-\left[A_{2}-\alpha_{1}\right]^{+}\right]^{+}, \nonumber 
\end{eqnarray}
\end{subequations}
which are obtained by summing up \eqref{UQ1}-\eqref{UQ4}.  In what follows, we analyze corner points of the DoF region and present the power allocation policy for $A_1,A_2$ case by case so that the DoF region in (7) is achieved. Below, the division of antenna configurations and CSIT qualities into general cases, and associated insights, follow that in \cite{hao2017achievable}.

\subsubsection{GENERAL CASE-1 (If $ \frac{N_2-N_1+(M-N_2)\alpha_{2}}{M-N_1} \leq \alpha_{1}$)} 
\begin{figure}[t]
\centering  %图片全局居中
%	\subfigbottomskip=2pt %两行子图之间的行间距
%	\subfigcapskip=-5pt %设置子图与子标题之间的距离
\subfigure[]{
\includegraphics[width=0.47\linewidth]{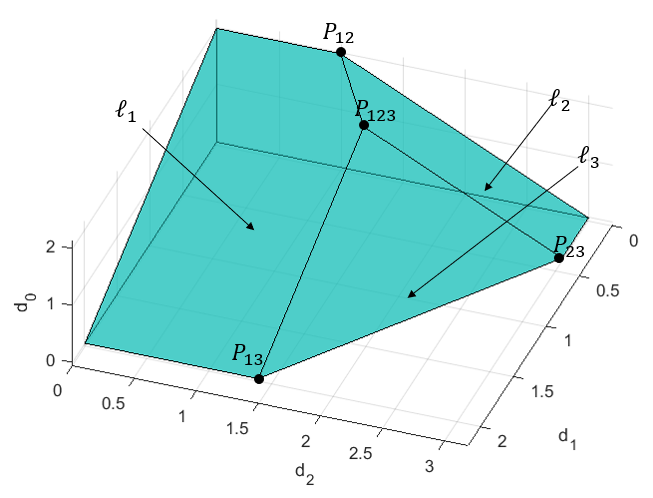} \label{F7}} 
\subfigure[]{
\includegraphics[width=0.47\linewidth]{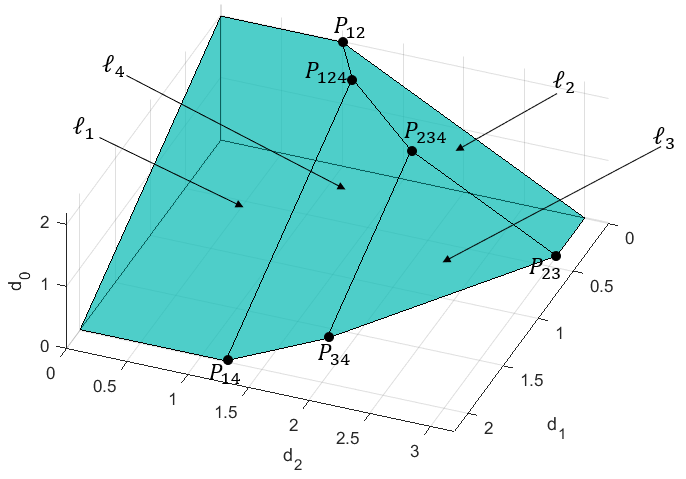}\label{F8}}
\\
\subfigure[]{
\includegraphics[width=0.47\linewidth]{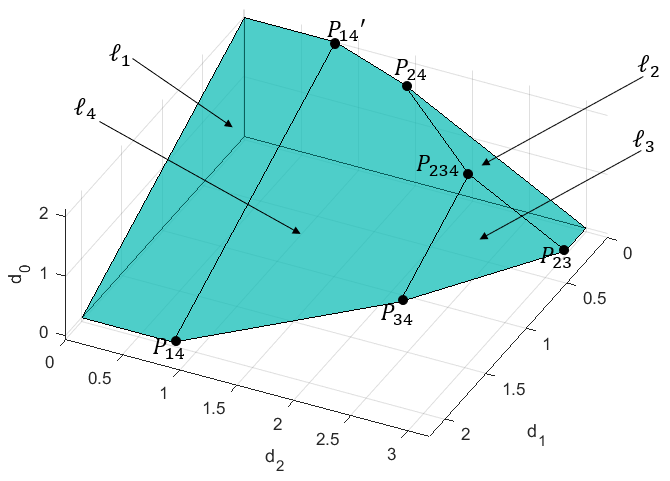}\label{F9}}
%\quad
\subfigure[]{
\includegraphics[width=0.47\linewidth]{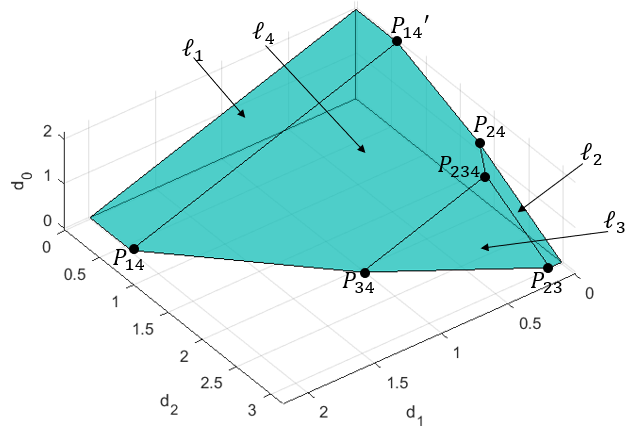}\label{F10}} 
\caption{ 
a) $\alpha_{1} \geq \frac{N_2-N_1+(M-N_2)\alpha_{2}}{M-N_1}$,
b) $\max\left\{1-\alpha_2,\frac{N_2-N_1}{M-N_1}\right\} \leq \alpha_{1} \leq \frac{N_2-N_1+(M-N_2)\alpha_{2}}{M-N_1}$,
c) $1-\alpha_{2} \leq \alpha_{1} \leq \frac{N_2-N_1}{M-N_1}$, and
d) $\alpha_{1} \leq 1-\alpha_{2}$ with only $\mathcal{P}_{234}$ exists.}
\end{figure}

The DoF region  is given below
\begin{eqnarray}\label{Ubound1}
\mathcal{D} =	\{	(d_1,d_2,d_0)  
\in \mathbb{R}_+^3| \qquad \qquad \qquad \qquad \qquad \quad \nonumber \\
\left.		 \begin{aligned}
& \ell_1: d_1+d_0 \le N_1, \\
& \ell_2: d_2+d_0 \le N_2, \\
& \ell_3: d_1+d_2+d_0 \le N_2+(M-N_2)\alpha_2,  \\
& \ell_4: \dfrac{d_1+d_0}{N_1} + \dfrac{d_2}{N_2} \leq 1+\dfrac{M-N_1}{N_2}\alpha_1. %\\
%\ell_5: \dfrac{d_1}{N_1} + \dfrac{d_2+d_0}{N_2} \leq 1 + \dfrac{M-N_1}{N_2}\alpha_1. \\	
\end{aligned}
\right\},    
\end{eqnarray} 
which is illustrated in Fig. \ref{F7}. The corner points on the coordinate are trivial and given by $\mathcal{P}_1 = (N_1, 0, 0)$, $\mathcal{P}_2=(0,N_2,0)$, $\mathcal{P}_0 = (0,0,N_1)$. The below proposition reveals the off-coordinate corner points.
%	\begin{figure}[t]
%		\centering
%		\includegraphics[width=8cm]{region_case1.png}
%		\caption{Achievable DoF region of $(M,N_1,N_2)$ MIMO BC.}\label{F7}
%	\end{figure}

\begin{proposition}
The off-coordinate corner points of the DoF region in GENERAL CASE-1 are given in the following. The strictly positive corner point is given by
\begin{itemize}
\item  $ \mathcal{P}_{123} =  
((M-N_2)\alpha_{2},  N_2-N_1+(M-N_2)\alpha_{2}, 
N_1-(M-N_2)\alpha_{2})$.
\end{itemize}	 
The other corner points with one zero coordinate are given by $\mathcal{P}_{13} = (N_1,N_2-N_1+(M-N_2)\alpha_{2},0),  \mathcal{P}_{23} = ((M-N_2)\alpha_{2},N_2,0)$, and $\mathcal{P}_{12} = (0,N_2-N_1,N_1)	
$. 
\end{proposition}
\begin{IEEEproof}
Please refer to Appendix A.
\end{IEEEproof}

Note that $\mathcal{P}_{ij}$ denotes the intersection of $\ell_i$ and $\ell_j$, and $\mathcal{P}_{ijk}$ denotes the intersection of $\ell_i$, $\ell_j$, and $\ell_k$.
Achieving the DoF region is equivalent to achieving corner points. We show that the above corner points are achievable by the following. In order to achieve the strictly positive corner point $\mathcal{P}_{123}$, we need to derive the power exponents $\left(A_{1}^{*}, A_{2}^{*}\right) := \arg \max d_{s}\left(A_{1}, A_{2}\right)$ that maximize the sum-DoF. As the sum-DoF does not change by adding a multicast message, according to \cite{hao2017achievable}, the optimal power exponents are given by 
\begin{equation}\label{UoptA1A2_1}
\begin{cases}
A_{1}^{*}=\alpha_{2}, \\  A_{2}^{*}= \max \left\{ \frac{N_2-N_1+(M-N_2)\alpha_{2}}{M-N_1},1-\frac{M-N_2}{N_2-N_1}\alpha_{1} \right\}.
\end{cases}
\end{equation}

By plugging $A_{1}^{*}$ and $A_{2}^{*}$ into \eqref{UQ1}-\eqref{UQ4} and considering no common part split from the unicast messages, i.e., $d_c=0$, we obtain $d_1=(M-N_2)\alpha_{2}$, $d_2=N_2-N_1+(M-N_2)\alpha_2$, and $d_0=N_1-(M-N_2)\alpha_2$, which is actually $\mathcal{P}_{123}$. Also, this implies a multicast message may mitigate the rate-splitting complexity of unicast messages. The achievability of corner points $\mathcal{P}_{13}$ and $\mathcal{P}_{23}$ are given in \cite{hao2017achievable} with $d_0 = 0$. To achieve corner point $\mathcal{P}_{12}$, substituting $A_1=0$ and $A_2=\frac{N_2-N_1}{M-N_1}$ into \eqref{UQ1}-\eqref{UQ4} and  assuming no common part split from the  unicast messages, i.e., $d_c=0$, we obtain $d_{1}=0$, $d_{2}=N_2-N_1$, $d_{0}=N_1$, which is actually $\mathcal{P}_{12}$.

\subsubsection{GENERAL CASE-2 (If $\max\left\{1-\alpha_2,\frac{N_2-N_1}{M-N_1}\right\} \leq \alpha_{1} \leq \frac{N_2-N_1+(M-N_2)\alpha_{2}}{M-N_1}$)}
The DoF region is given below
\begin{eqnarray}\label{Ubound2} 
\mathcal{D} = 	\left\{(d_1,d_2,d_0)  
\in \mathbb{R}_+^3|	\ell_1,\ell_2,\ell_4 \,\text{in (11)},   \qquad \qquad  \qquad \,\,\,\nonumber \right. \\ \left. 
\begin{aligned}
&  \ell_3: d_1+d_2+d_0 \le N_2 + \\
&    (M-N_2)\left(-\frac{N_2-N_1}{M-N_1}+\frac{N_2-N_1}{M-N_1}\alpha_{2}+\alpha_{1}\right).  
\end{aligned}
\right\},  
\end{eqnarray}		
which is illustrated in Fig. \ref{F8}. The corner points on the coordinate are trivial and given by $\mathcal{P}_1 = (N_1, 0, 0)$, $\mathcal{P}_2=(0,N_2,0)$, $\mathcal{P}_0 = (0,0,N_1)$. The below proposition reveals the off-coordinate corner points.

\begin{proposition}
The off-coordinate corner points of the DoF region in GENERAL CASE-2 are given in the following. The strictly positive corner points are given by
\begin{itemize}
\item	$\mathcal{P}_{234}=\frac{1}{M-N_1}((M-N_2)(N_1-N_2+(M-N_1)\alpha_1 + (N_2-N_1)\alpha_2),(M^2+N_1^2-2MN_1-MN_2+N_1N_2)\alpha_1 + (MN_2-N_2^2)\alpha_2 + N_2^2 - N_1N_2, (2MN_1+MN_2 - N_1N_2 -M^2 -N_1^2)\alpha_1 + (N_2^2 - MN_2)\alpha_2 + MN_2-N_2^2)$,   \item $\mathcal{P}_{124}=((M-N_1)\alpha_{1}-(N_2-N_1), (M-N_1)\alpha_1,N_2-(M-N_1)\alpha_{1})$.
\end{itemize} 
The other corner points with one coordinate is zero are given by $\mathcal{P}_{34}=(N_1\alpha_1-\frac{N_1(M-N_2)}{M-N_1}\alpha_{2}+\frac{N_1(M-N_2)}{M-N_1},   -(N_1$ $+N_2-M)\alpha_{1} + \frac{N_2(M-N_2)}{M-N_1}\alpha_{2}   +\frac{N_2(N_2-N_1)}{M-N_1},0), 	\mathcal{P}_{23}=((M-N_2)\alpha_{1}+\frac{(M-N_2)(N_2-N_1)}{M-N_1}\alpha_{2}-\frac{(M-N_2)(N_2-N_1)}{M-N_1}$ $,N_2,0), 	
\mathcal{P}_{14}=(N_1,(M-N_1)\alpha_{1},0)$, and 	
$\mathcal{P}_{12}=(0,N_2-N_1,N_1). %\\
%\mathcal{P}_1 = (), \\
%\mathcal{P}_2 = (), \\
%\mathcal{P}_0 = (), 	
$

\end{proposition}
\begin{IEEEproof}
Please refer to Appendix A.
\end{IEEEproof} 

We show that the above corner points are achievable by the following. It can be checked that $(A_1,A_2)$ to achieve corner points $\mathcal{P}_{12}$, $\mathcal{P}_{234}$, $\mathcal{P}_{34}$ and $\mathcal{P}_{23}$ are the same as $(A_1,A_2)$ to achieve corner points $\mathcal{P}_{12}$, $\mathcal{P}_{123}$, $\mathcal{P}_{13}$ and $\mathcal{P}_{23}$ in GENERAL CASE-1. In order to achieve corner points $\mathcal{P}_{124}$ and $\mathcal{P}_{14}$, we substitute  $A_1=\frac{(M-N_1)\alpha_{1}-(N_2-N_1)}{M-N_2}$ and $A_2=\alpha_{1}$ into \eqref{UQ1}-\eqref{UQ4} and yields  $d_{p1}=(M-N_1)\alpha_{1}-(N_2-N_1)$, $d_{p2}=(M-N_1)\alpha_{1}$ and $d_c+d_0=N_2-(M-N_1)\alpha_{1}$. If there is no common part split from the unicast messages, i.e., $d_c=0$, we obtain $d_1=d_{p1}=(M-N_1)\alpha_{1}-(N_2-N_1)$, $d_2=d_{p2}=(M-N_1)\alpha_{1}$, $d_0=N_2-(M-N_1)\alpha_{1}$, which is actually $\mathcal{P}_{124}$. If there is no multicast message, i.e. $d_0=0$, and the multicast message only carries information intended for user 1, i.e. $d_1=d_{p1}+d_c$, we obtain $d_1=N_1$, $d_2=(M-N_1)\alpha_1$, and $d_0=0$, which is actually $\mathcal{P}_{14}$.

%\begin{figure}[t]    % Fig. \ref{F8}
%	\centering
%	\includegraphics[width=8cm]{region_case2A.png}
%	\caption{Achievable DoF region of $(M,N_1,N_2)$ MIMO BC with $\max\left\{1-\alpha_2,\frac{N_2-N_1}{M-N_1}\right\} \leq \alpha_{1} \leq \frac{N_2-N_1+(M-N_2)\alpha_{2}}{M-N_1}$.} \label{F8}
%\end{figure}
%\begin{figure}[t]   % Fig. \ref{F9}
%	\centering
%	\includegraphics[width=8cm]{region_case2B.png}
%	\caption{Achievable DoF region of $(M,N_1,N_2)$ MIMO BC with $ 1-\alpha_2 \leq \alpha_{1} \leq \frac{N_2-N_1}{M-N_1}$.} \label{F9}
%\end{figure}
\subsubsection{GENERAL CASE-3 (If $1-\alpha_{2} \leq \alpha_{1} \leq \frac{N_2-N_1}{M-N_1}$)}

The DoF region of this case is the same as that in GENERAL CASE-2  except a different shape due to different CSIT qualities, where this DoF region is illustrated in Fig. \ref{F9}.
Thereby, corner points on the coordinate are trivial and given by $\mathcal{P}_1 = (N_1, 0, 0)$, $\mathcal{P}_2=(0,N_2,0)$, $\mathcal{P}_0 = (0,0,N_1)$. The below proposition reveals the off-coordinate corner points.
\begin{proposition}
The off-coordinate corner points of the DoF region in GENERAL CASE-3 are given by two groups. The first group includes corner points $\mathcal{P}_{14}'=(0,(M-N_1)\alpha_{1},N_1)$, and
$\mathcal{P}_{24}=(0,N_2-\frac{(M-N_1)N_1}{N_2-N_1}\alpha_{1},\frac{(M-N_1)N_1}{N_2-N_1}\alpha_{1})$. The second group includes corner points $\mathcal{P}_{234}$, $\mathcal{P}_{34}$, $\mathcal{P}_{23}$ and $\mathcal{P}_{14}$, which are the same as that in GENERAL CASE-2.  
\end{proposition}

\begin{IEEEproof}
For $d_1 = 0$, it turns out that $\mathcal{D} = \{(d_2,d_0)\in \mathbb{R}_+^2|d_0\le N_1, d_2+d_0 \le N_2, \frac{d_0}{N_1} + \frac{d_2}{N_2} \le 1 + \frac{M-N_1}{N_2}\alpha_1 \}$. The off-coordinate corner points are given by $\mathcal{P}_{14}'= (0,(M-N_1)\alpha_{1},N_1)$ and
$\mathcal{P}_{24}=(0,N_2-\frac{(M-N_1)N_1}{N_2-N_1}\alpha_{1},\frac{(M-N_1)N_1}{N_2-N_1}\alpha_{1})$. The derivations of remaining corner points are the same as those in Proposition 2.
\end{IEEEproof}

We show that the corner points in the first group are achievable by the following. In order to achieve corner point $\mathcal{P}_{14}'$, we substitute $A_1=0$ and $A_2=1-\alpha_{1}$ into \eqref{UQ1}-\eqref{UQ4}, and consider there is no common part split from the unicast messages. Then we obtain $d_{1}=0$, $d_{2}=(M-N_1)\alpha_{1}$ and $d_0=N_1$, which is actually $\mathcal{P}_{14}'$. To achieve corner point $\mathcal{P}_{24}$, we substitute $A_1=0$ and $A_2=1-\frac{M-N_2}{N_2-N_1}\alpha_{1}$ into \eqref{UQ1}-\eqref{UQ4}, and consider there is no common part split from the unicast messages. Then, we obtain $d_{1}=0$, $d_{2}=N_2-\frac{(M-N_1)N_1}{N_2-N_1}\alpha_{1}$, and $d_{0}=\frac{(M-N_1)N_1}{N_2-N_1}\alpha_{1}$, which is actually $\mathcal{P}_{24}$.

\subsubsection{GENERAL CASE-4 (If $  \alpha_{1} \leq 1-\alpha_{2}$)}
The DoF region is given below
\begin{eqnarray}\label{Ubound4}
\mathcal{D} = 	\left\{(d_1,d_2,d_0)  
\in \mathbb{R}_+^3|	\ell_1,\ell_2,\ell_4 \,\text{in (11)},   \qquad \qquad  \qquad \,\,\,\nonumber \right. \\ \left. 
\begin{aligned}
&  \ell_3: d_1+d_2+d_0 \le N_2 + \\
&    \frac{\alpha_{1}\alpha_{2}(M-N_2)^2}{(N_2-N_1)(1-\alpha_{1})+(M-N_2)\alpha_{2}}.  
\end{aligned}
\right\},  
\end{eqnarray}
which is illustrated in Fig. \ref{F10}. The corner points on the coordinate are trivial and given by $\mathcal{P}_1 = (N_1, 0, 0)$, $\mathcal{P}_2=(0,N_2,0)$, $\mathcal{P}_0 = (0,0,N_1)$. The below proposition reveals the off-coordinate corner points of the DoF region.

\begin{proposition}
The off-coordinate corner points of the DoF region in GENERAL CASE-4 are given by two groups. The first group includes the following strictly positive corner points:
\begin{itemize}
\item 	$
\mathcal{P}_{234}=\frac{1}{\Delta}((M-N_2)^2\alpha_1\alpha_2,$ $
\left(M-N_{1}\right)N_1\alpha_1^2+( (M-N_2)(M-N_1-N_2)\alpha_2 + N_1^2 - N_2^2 - MN_1 + N_1N_2)\alpha_1 + (M-N_2)N_2\alpha_2 + N_2^2 - N_1N_2$, 	$((M-N_2)(N_1+N_2-M)\alpha_2 + (M-N_1)N_1(1-\alpha_1))\alpha_1)$ if $\alpha_1 \le \frac{N_2-N_1}{M-N_1} + \frac{M-N_2}{M-N_1}\alpha_2$,
\item 	$\mathcal{P}_{123} = (\frac{(M-N_2)^2\alpha_1\alpha_2}\Delta,N_2-N_1 + \frac{(M-N_2)^2\alpha_1\alpha_2}\Delta, N_1 - \frac{(M-N_2)^2\alpha_1\alpha_2}\Delta)$ if $\frac{(M-N_2)^2\alpha_1\alpha_2}\Delta \le \min\{(M-N_1)\alpha_1 - (N_2-N_1), N_1\}$,
\item $\mathcal{P}_{124} = ((M-N_1)\alpha_1 - (N_2 - N_1), (M-N_1)\alpha_1, N_2 - (M-N_1)\alpha_1)$ if 	$(M-N_1)\alpha_1 - (N_2-N_1) \le \frac{(M-N_2)^2\alpha_1\alpha_2}\Delta$ and $\frac{N_2-N_1}{M-N_1} \le \alpha_1 \le \frac{N_2}{M-N_1}$.
\end{itemize}  	
The other corner points in this group with one coordinate is zero are given by
$\mathcal{P}_{34}= \frac{1}{\Delta}(N_1\alpha_1(M-N_1-(M-N_1)\alpha_1 + (M-N_2)\alpha_2) 
,
\left(M-N_{1}\right)N_1\alpha_1^2+( (M-N_2)(M-N_1-N_2)\alpha_2 + N_1^2 - N_2^2 - MN_1 + N_1N_2)\alpha_1 + (M-N_2)N_2\alpha_2 + N_2^2 - N_1N_2,  0)$, 		
$\mathcal{P}_{23}=(\frac{(M-N_2)^2\alpha_1\alpha_2}{\Delta}, N_2, 0),$
where $\Delta = (M-N_2)\alpha_2 + (N_2 - N_1)(1-\alpha_1)$.		

The second group includes $\mathcal{P}_{14}$, $\mathcal{P}_{14}'$, $\mathcal{P}_{24}$, which are the same as that in GENERAL CASE-3.   
\end{proposition}
\begin{IEEEproof}
Please refer to Appendix A.
\end{IEEEproof}

We show that the above corner points in the first group are achievable by the following. Different from other cases, the space-time rate-splitting transmission scheme provided by \cite{hao2017achievable} is applied to further increase the sum-DoF in this case. Suppose there are totally $T$ time slots during the transmission, and $T$ is sufficiently large. The power exponents in time slot $l$, denoted by $(A_{1,l},A_{2,l})$, are given by	
\begin{equation}
(A_{1,l},A_{2,l})= 
\begin{cases} 
(\alpha_2,1),         & {l = 1,...,\rho T},\\
(\alpha_{2},\alpha_{1}),      & {l = \rho T + 1,...,T},\\
\end{cases}   \label{PowerEx}
\end{equation} 
where $\rho \in [0,1]$. According to \cite{hao2017achievable},  optimal $\rho^*$ is  given by
\begin{equation}
\rho^*=\dfrac{N_2-N_1-(M-N_1)\alpha_{1}+(M-N_2)\alpha_{2}}{(N_2-N_1)(1-\alpha_{1})+(M-N_2)\alpha_{2}}.
\end{equation}
Therefore, this space-time transmission scheme can achieve
\begin{subequations}
\begin{eqnarray}
&& \!\!\!\!\!\!\!\!\!\!\!\!	d_{p1} \le (M-N_2)(\rho^*[\alpha_1+\alpha_2-1]^+(1-\rho^*)\alpha_2), \\
&&\!\!\!\!\!\!\!\!\!\!\!\!	d_{p2} \le \rho^*(N_2-(N_1+N_2-M)\alpha_1) + \nonumber \\
&&  \qquad \qquad \qquad \qquad  (1-\rho^*)(M-N_1)\alpha_1, \\
&&\!\!\!\!\!\!\!\!\!\!\!\!	d_c + d_0 \le \rho^*(N_1+N_2-M)\alpha_1  + \nonumber \\
&& \qquad \qquad \qquad  (1-\rho^*)(N_2 - (M-N_1)\alpha_1).
\end{eqnarray}
\end{subequations}

% Adapting power allocation across the $T$ slots and considering there is no common part split from unicast message}s, we can compute $d_1=\frac{1}{\Delta}((M-N_2)^2\alpha_1\alpha_2$, $d_2=\left[M-N_{1}\right)N_1\alpha_1^2+( (M-N_2)(M-N_1-N_2)\alpha_2 + N_1^2 - N_2^2 - MN_1 + N_1N_2)\alpha_1 + (M-N_2)N_2\alpha_2 + N_2^2 - N_1N_2$, and $d_0=((M-N_2)(N_1+N_2-M)\alpha_2 + (M-N_1)N_1(1-\alpha_1))\alpha_1)$ if $\alpha_1 \le \frac{N_2-N_1}{M-N_1} + \frac{M-N_2}{M-N_1}\alpha_2$, which is actually $\mathcal{P}_{234}$. Similarly, $\mathcal{P}_{123}$ and $\mathcal{P}_{124}$ can be achieved under corresponding condition of CSIT quality. $(A_1,A_2)$ to achieve corner points $\mathcal{P}_{34}$ and $\mathcal{P}_{23}$ are given in \cite{hao2017achievable} with $d_0$ set to zero. 

The space-time transmission is adapted to achieve corner points $\mathcal{P}_{234}$, $\mathcal{P}_{123}$, $\mathcal{P}_{34}$ and $\mathcal{P}_{23}$. The power exponents $(A_{1,l},A_{2,l})$ to achieve corner points $\mathcal{P}_{34}$ and $\mathcal{P}_{23}$ are given in \cite{hao2017achievable} with $d_0 = 0$. In particular, to achieve corner point $\mathcal{P}_{234}$, we consider there is no common part split from the unicast messages, i.e., $d_c=0$, and $d_0 = \frac{1}{\Delta}((M-N_2)(N_1+N_2-M)\alpha_2 + (M-N_1)N_1(1-\alpha_1)\alpha_1)$, then corner point $\mathcal{P}_{234}$ is achieved.  To achieve corner point $\mathcal{P}_{123}$, we consider there is no common part split from the unicast messages, i.e., $d_c=0$, $d_{p1} = \frac{(M-N_2)^2\alpha_1\alpha_2}{\Delta}$, $d_0 = N_1-\frac{(M-N_2)^2\alpha_1\alpha_2}{\Delta}$, then corner point $\mathcal{P}_{123}$ is achieved. The corner points $\mathcal{P}_{14}$, $\mathcal{P}_{14}'$ and $\mathcal{P}_{24}$ are the same as CASE-3 with the same solutions to achieve. The corner point $\mathcal{P}_{124}$ is the same as CASE-2, thus the achievability can also be proved.

%	It worth mentioning that if we do not apply space-time transmission, the achievable sum-DoF would be $(M-N_2)\alpha_{1}+\frac{(M-N_2)(N_2-N_1)}{M-N_1}(\alpha_{2}-1)+N_2$.

%{\color{blue}{\subsubsection{GENERAL CASE-5 (When $\alpha_{1} \leq  \left\{1-\alpha_2,\frac{N_2-N_1}{M-N_1}\right\}$)}
%	The DoF region in this case is the same as that in GENERAL CASE-4 with the same solution to achieve corner points.}}

\subsection{An Example by Two-User $(4, 2, 3)$ MIMO BC}
\begin{figure}
\centering
\includegraphics[width=3.25in]{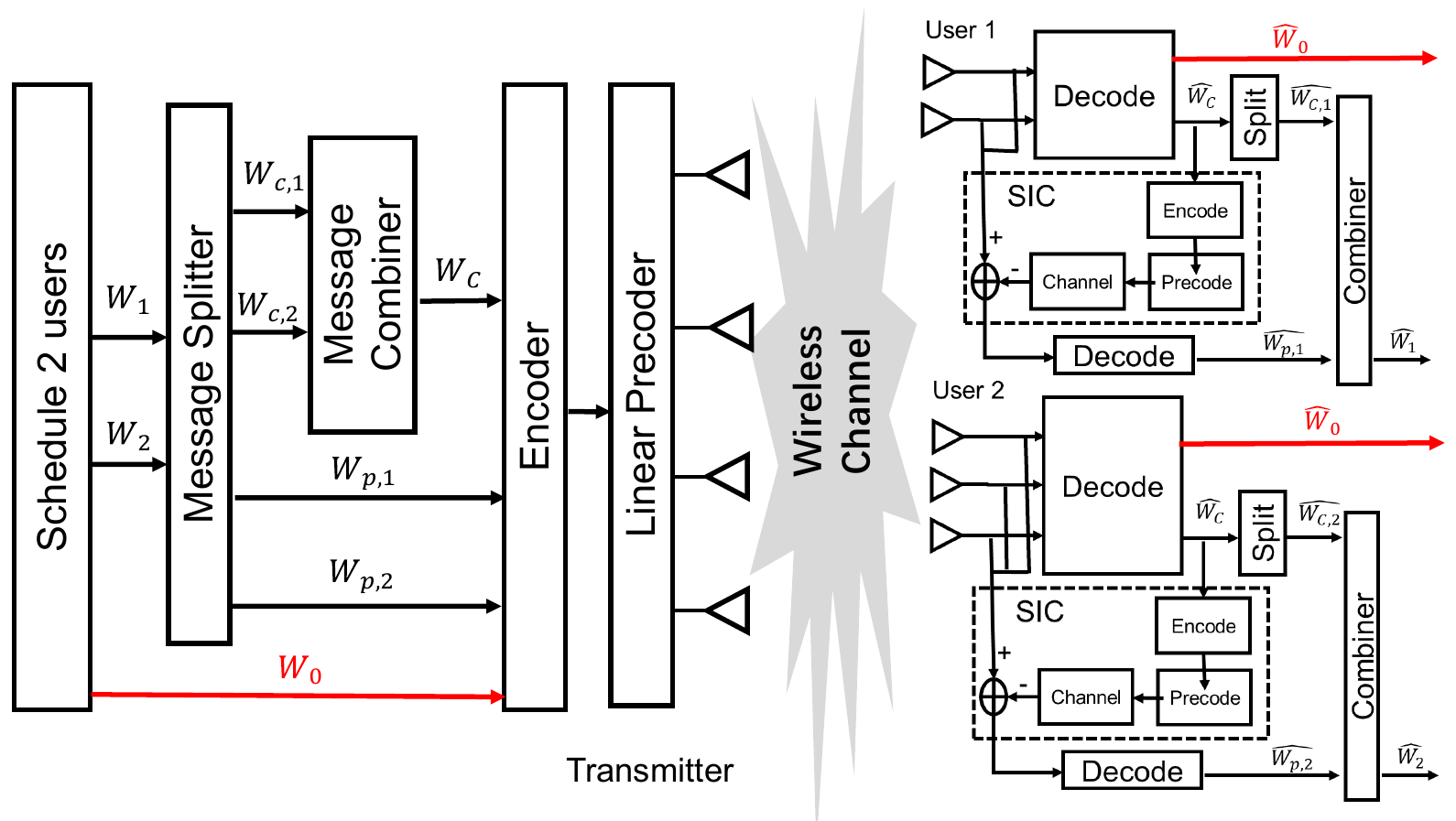}
\caption{System diagram of this $(4,2,3)$ hybrid message two-user MIMO BC.} \label{Dia}
\end{figure}
In this subsection, to provide better understanding, we consider a special case of a MIMO system with $(M,N_1,N_2)=(4,2,3)$. Note that with the same principle, other examples such as MIMO BC $(4,1,3)$ and MIMO BC $(4,2,1)$ are left for readers. The system diagram of this $(4,2,3)$ hybrid message MIMO system is illustrated in Fig. \ref{Dia}, where the main difference from conventional rate-splitting system with unicast messages is highlighted in red color.
We constitute the rate-splitting transmission block for the two-user $(4, 2, 3)$ MIMO BC with hybrid messages and imperfect CSIT as follows. 
\begin{itemize}
\item $1$ unicast symbol, denoted by $u_1$, is transmitted to $\text{Rx}_1$ along a ZF-precoder $\mathbf{v}_1=\widehat{\textbf{H}}^{\perp}_2 \in \mathbb{C}^{4 \times 1}$ with power exponent $A_1$;
\item $2$ unicast symbols, denoted by $\mathbf{u}_2^{1} \in \mathbb{C}^{2 \times 1}$, are transmitted to $\text{Rx}_2$ along a ZF-precoder $\mathbf{V}_2^{1}=\widehat{\textbf{H}}_1^{\perp} \in \mathbb{C}^{4 \times 2}$ with power exponent $A_2$;
\item $1$ unicast symbol, denoted by  $u_2^{2}$, is transmitted to $\text{Rx}_2$ along a precoder $\mathbf{v}_2^{2} \in \mathbb{C}^{4 \times 1}$ in the subspace spanned by $\widehat{\textbf{H}}_2$. Its power exponent is $[A_2-\alpha_1]^+$;
\item The composite multicast symbol, denoted by $\mathbf{(c+u_0)} \in \mathbb{C}^{4 \times 1}$, is transmitted with remaining power.
\end{itemize}
Recalling that the power exponents $A_1$ and $A_2$ are restricted in $A_1\in[0,\alpha_2]$ and $A_2\in[0,1]$. Mathematically, the transmitted and received signals can be written as 
\begin{subequations}
\begin{equation}
\mathbf{s}=\underbrace{\mathbf{c+u_0}}_{\sim P}+\underbrace{\mathbf{v}_{1} u_{1}}_{\sim P^{A_{1}}}+\underbrace{\mathbf{V}_{2}^{1} \mathbf{u}_{2}^{1}}_{\sim P^{A_{2}}}+\underbrace{\mathbf{v}_{2}^{2} u_{2}^{2}}_{\sim P^{\left[A_{2}-\alpha_{1}\right]^{+}}}, 
\end{equation}
\begin{equation}
\mathbf{y}_{1}=\underbrace{\mathbf{H}_{1}^{H} \mathbf{(c+u_0)}}_{\sim P} + \underbrace{\mathbf{H}_{1}^{H} \mathbf{v}_{1} u_{1}}_{\sim P^{A_{1}}}+\underbrace{\mathbf{H}_{1}^{H}\left(\mathbf{V}_{2}^{1} \mathbf{u}_{2}^{1}+\mathbf{v}_{2}^{2} u_{2}^{2}\right)}_{\sim P^{[\left.A_{2}-a_{1}\right]^{+}}}, 
\end{equation}
\begin{equation}
\mathbf{y}_{2}=\underbrace{\mathbf{H}_{2}^{H} \mathbf{(c+u_0)}}_{\sim P} + \underbrace{\mathbf{H}_{2}^{H} \mathbf{v}_{1} u_{1}}_{\sim P^{A_1-\alpha_{2}}}+\underbrace{\mathbf{H}_{2}^{H} \mathbf{V}_{2}^{1} \mathbf{u}_{2}^{1}}_{\sim P^{A_{2}}}+\underbrace{\mathbf{H}_{2}^{H} \mathbf{v}_{2}^{2} u_{2}^{2}}_{\sim P^{\left[A_{2}-\alpha_{1}\right]^+}} .
\end{equation}
\end{subequations}

% As we can see from the received signals, if $A_2 \leq \alpha_1$, the undesired unicast symbols are dissolved into the noise. If $A_2 \textgreater \alpha_1$, the designed power allocation policy will ensure that all the three unicast symbols intended for $\text{Rx}_2$ are received by $\text{Rx}_1$ with the same power level.  
Similar to \cite{hao2017achievable}, for (18b) and (18c), using the proof in \cite[Appendix A]{hao2017achievable} and considering that each receiver decodes the common part split from the unicast messages and the multicast message successively, the following DoF tuple is achievable.

At $\text{Rx}_1$, we have 
\begin{subequations}
\begin{equation}\label{Q1}
d_{0}+d_{c} \leq d_c^1 := 2-\max \left\{A_{1}, A_{2}-\alpha_{1}\right\}-\left[A_{2}-\alpha_{1}\right]^{+}  ,
\end{equation}
\begin{equation}\label{Q2}
d_{p1} \le \left[A_{1}-\left[A_{2}-\alpha_{1}\right]^{+}\right]^{+} ,
\end{equation}

At $\text{Rx}_2$, we have 
\begin{equation}\label{Q3}
d_{0}+d_{c} \leq d_c^2 := 3-2 A_{2}-\left[A_{2}-\alpha_{1}\right]^{+},
\end{equation}
\begin{equation}\label{Q4}
d_{p 2} \le 2 A_{2}+\left[A_{2}-\alpha_{1}\right]^{+}.
\end{equation}
\end{subequations}

Accordingly, the achievable sum-DoF is defined as 
\begin{equation}\label{sumDoF}
d_{s}(A_1,A_2):= \min\{d_s^{1}(A_2),d_s^{2}(A_1,A_2)\},
\end{equation}
where
\begin{subequations}
\begin{equation}\label{sumDoF1}
d_s^1\left(A_{2}\right)=2+2 A_{2}-\left[A_{2}-\alpha_{1}\right]^{+}, \nonumber 
\end{equation}
\begin{equation}\label{sumDoF2}
d_s^2\left(A_{1}, A_{2}\right)=3+\left[A_{1}-\left[A_{2}-\alpha_{1}\right]^{+}\right]^{+}, \nonumber
\end{equation}
\end{subequations}
which are obtained by summing up \eqref{Q1}-\eqref{Q4}. In the following, we present the power allocation policy case by case so that the DoF region can be achieved.

\subsubsection{CASE-1 (If $ \frac{1+\alpha_{2}}{2} \leq \alpha_{1}$)}
Via GENERAL CASE-1, the DoF region of this $(4, 2, 3)$ MIMO BC with hybrid messages and imperfect CSIT is simplified to
\begin{eqnarray}\label{bound1}
\mathcal{D} =	\{	(d_1,d_2,d_0)  
\in \mathbb{R}_+^3| \qquad \qquad \qquad \quad  \nonumber \\
\left.		 \begin{aligned}
&	\ell_1: d_1+d_0 \le 2, \\
& \ell_2: d_2+d_0 \le 3, \\ 	
&	\ell_4: d_1+d_2+d_0 \le 3 + \alpha_2,  \\  
&\ell_4: \dfrac{d_1}{2} + \dfrac{d_2}{3} + \dfrac{d_0}{2}\leq 1 + \dfrac{2}{3}\alpha_1. \\	
\end{aligned}
\right\}.    
\end{eqnarray}
% \begin{figure}[t]
% 	\centering
% 	\includegraphics[width=8cm]{region_case1.png}
% 	\caption{Achievable DoF region of $(M,N_1,N_2)$ MIMO BC with $\Phi \le 0$.}\label{F3}
% \end{figure} 

%\begin{figure}[t]
%	\centering  %图片全局居中
%	\subfigbottomskip=2pt %两行子图之间的行间距
%	\subfigcapskip=-5pt %设置子图与子标题之间的距离
%	\subfigure[]{
%		\includegraphics[width=0.465\linewidth]{region_case1.png} \label{F3}} 
%	\subfigure[]{
%		\includegraphics[width=0.465\linewidth]{region_case2A.png}\label{F4}}
%	\\
%	\subfigure[]{
%		\includegraphics[width=0.465\linewidth]{region_case2B.png}\label{F5}}
%\quad
%	\subfigure[]{
%		\includegraphics[width=0.465\linewidth]{region_case3.png}\label{F6}} 
%	\caption{Achievable DoF region of $(M,N_1,N_2)$ MIMO BC with 
%		a) $\Phi \le 0$,
%		b) $\max\{1-\alpha_2,\frac{1}{2}\} \leq \alpha_{1} \leq \frac{1+\alpha_{2}}{2}$,
%		c) $1-\alpha_2 \leq \alpha_{1} \leq \frac{1}{2}$, and
%		d) $\frac{1-\alpha_{2}}{2} \leq \alpha_{1} \leq 1-\alpha_2$}.
%\end{figure}

The on-coordinate corner points are given by $(2,0,0)$, $(0,3,0)$, and $(0,0,2)$, which can be trivially achieved. According to Proposition 1, all off-coordinate corner points are summarized as
$\mathcal{P}_{123} = (\alpha_{2},1+\alpha_{2},2-\alpha_{2})$,  
$\mathcal{P}_{13} = (2,1+\alpha_{2},0)$,  
$\mathcal{P}_{23} = (\alpha_{2},3,0)$,  and 
$\mathcal{P}_{12} = (0,1,2)$. 
We then show that the above off-coordinate corner points are achievable.

To achieve corner points $\mathcal{P}_{123}$, $\mathcal{P}_{13}$, and $\mathcal{P}_{23}$, we need to derive the optimal power exponents $A_1^*$ and $A_2^*$. As shown in \cite{hao2017achievable}, $\left(A_{1}^{*}, A_{2}^{*}\right) := \arg \max d_{s}\left(A_{1}, A_{2}\right)$ are given by 
\begin{equation}\label{optA1A2_1}
\begin{cases}
A_{1}^{*}=\alpha_{2}, \\ A_{2}^{*}=\max \left\{ \frac{1+\alpha_{2}}{2},1-\alpha_{1} \right\}.
\end{cases}	
\end{equation}
Hence, to achieve corner point $\mathcal{P}_{123}$, we substitute \eqref{optA1A2_1} into \eqref{Q1}-\eqref{Q4}, and consider no common part split from the unicast messages, i.e., $d_c=0$. Then, we have $d_1=\alpha_2$, $d_2=1+\alpha_{2}$, and $d_0= 2-\alpha_{2}$, which is actually $\mathcal{P}_{123}$. The achievability to achieve corner points $\mathcal{P}_{13}$ and $\mathcal{P}_{23}$ are presented in \cite{hao2017achievable} with $d_0 = 0$. To achieve corner point $\mathcal{P}_{12}$, substituting $A_1=0$ and $A_2=\frac{1}{2}$ into \eqref{Q1}-\eqref{Q4} yields $d_{p1}=0$, $d_{p2}=1$, $d_{c}+d_{0}=2$. If no common part is split from the unicast messages, i.e., $d_c=0$, with $A_1=0,A_2=\frac{1}{2}$, corner point $\mathcal{P}_{12}=(0,1,2)$ can be achieved.

\subsubsection{CASE-2 (If $ \max\{1-\alpha_2,\frac{1}{2}\} \leq \alpha_{1} \leq \frac{1+\alpha_{2}}{2}$)}
Via GENERAL CASE-2, the DoF region of this $(4, 2, 3)$ MIMO BC with imperfect CSIT and hybrid message is simplified to	
\begin{eqnarray}\label{bound2}
\mathcal{D} =	\{	(d_1,d_2,d_0)  
\in \mathbb{R}_+^3| \qquad \qquad \qquad \qquad \qquad \quad \nonumber \\
\left.		 \begin{aligned}
& \ell_1,\ell_2,\ell_4 \text{ in } \eqref{bound1}, \\
& \ell_3: d_1+d_2+d_0 \le 3 +\dfrac{-1+2\alpha_{1}+\alpha_{2}}{2}.  
%\ell_5: \dfrac{d_1}{N_1} + \dfrac{d_2+d_0}{N_2} \leq 1 + \dfrac{M-N_1}{N_2}\alpha_1. \\	
\end{aligned}
\right\}.   
\end{eqnarray}  
% \begin{figure}[t]
% 	\centering
% 	\includegraphics[width=8cm]{region_case2A.png}
% 	\caption{Achievable DoF region of $(M,N_1,N_2)$ MIMO BC with $ \max\{1-\alpha_2,\frac{1}{2}\} \leq \alpha_{1} \leq \frac{1+\alpha_{2}}{2}$.} \label{F4}
% \end{figure}

The on-coordinate corner points are given by $(2,0,0)$, $(0,3,0)$, and $(0,0,2)$, which can be trivially achieved. 
According to Proposition 2,  all off-coordinate corner points are summarized as
$\mathcal{P}_{124}=(2\alpha_{1}-1, 2\alpha_{1}, 3-2\alpha_{1})$,   $\mathcal{P}_{234}=(\alpha_1+\frac{1}{2}\alpha_2-\frac{1}{2}, \frac{3}{2}\alpha_2-\alpha_1+\frac{3}{2}, \alpha_1-\frac{3}{2}\alpha_{2}+\frac{3}{2})$, 
$\mathcal{P}_{14}=(2,2\alpha_1,0)$,  $\mathcal{P}_{23}=\left(\frac{-1+2\alpha_1+\alpha_2}{2},3,0\right),  
\mathcal{P}_{12}=(0,1,2)$, and 
$\mathcal{P}_{34}=\left(2\alpha_1-\alpha_2+1 \right.$, $\left. \frac{3}{2}\alpha_2-\alpha_1+\frac{3}{2},0\right)$. 
We next show that the above off-coordinate corner points are achievable.

To achieve corner point $\mathcal{P}_{124}$, substituting $A_1=2\alpha_{1}-1$ and $A_2=\alpha_{1}$ into \eqref{Q1}-\eqref{Q4} yields $d_{p1}=2\alpha_{1}-1$, $d_{p2}=2\alpha_{1}$ and $d_c+d_0=3-2\alpha_{1}$. If there is no common part split from the unicast messages, i.e., $d_c=0$, we have $d_{1}=2\alpha_{1}-1$, $d_{p2}=2\alpha_{1}$ and $d_0=3-2\alpha_{1}$, which is actually $\mathcal{P}_{124}$. The solutions to achieve corner point  $\mathcal{P}_{234}$ is the same as that achieving corner point $\mathcal{P}_{123}$ in CASE-1. To achieve corner points $\mathcal{P}_{14}$, substituting $A_1=2\alpha_{1}-1$ and $A_2=\alpha_{1}$ into \eqref{Q1}-\eqref{Q4} yields $d_{p1}=2\alpha_{1}-1$, $d_{p2}=2\alpha_{1}$ and $d_c+d_0=3-2\alpha_{1}$. If there is no multicast message, i.e., $d_0=0$, and the common part split from the unicast messages, then $d_{1}=2$, $d_{2}=2\alpha_{1}$ and $d_0=0$, which is actually $\mathcal{P}_{14}=(2,2\alpha_1,0)$. $(A_1,A_2)$ to achieve corner points $\mathcal{P}_{12}$, $\mathcal{P}_{34}$ and $\mathcal{P}_{23}$ are the same as that to achieve corner points $\mathcal{P}_{12}$, $\mathcal{P}_{13}$ and $\mathcal{P}_{23}$ in CASE-1.

\subsubsection{CASE-3 (If $ 1-\alpha_2 \leq \alpha_{1} \leq \frac{1}{2}$)}
The  inequalities of this DoF region is the same as those in CASE-2. The unique difference is the region shape due to different CSIT qualities.   The on-coordinate corner points are given by $(2,0,0)$, $(0,3,0)$, and $(0,0,2)$, which can be trivially achieved.  According to Proposition 3, all off-coordinate corner points are given by
$
\mathcal{P}_{234}=\left(\alpha_1+\frac{\alpha_2}{2}-\frac{1}{2}, \frac{3\alpha_2}{2}-\alpha_1+\frac{3}{2}, \alpha_1-\frac{3\alpha_{2}}{2}+\frac{3}{2}\right), 
\mathcal{P}_{14}=(2,2\alpha_1,0),  
\mathcal{P}_{34}=(2\alpha_1-\alpha_2+1,\frac{3}{2}\alpha_2-\alpha_1+\frac{3}{2},0), 
\mathcal{P}_{23}=\left(\frac{-1+2\alpha_1+\alpha_2}{2},3,0\right),  
\mathcal{P}_{14}'=(0,2\alpha_1,2)$,  and
$\mathcal{P}_{24}=(0,3-4\alpha_{1},4\alpha_{1}).  
$ 
We next show that the above off-coordinate corner points are achievable. 

The corner points $\mathcal{P}_{234}$, $\mathcal{P}_{34}$, $\mathcal{P}_{23}$, and $\mathcal{P}_{14}$ are the same as that in CASE-2, thus with the same solutions to achieve.  
To achieve corner point $\mathcal{P}_{24}$, substituting $A_1=0$ and $A_2=1-\alpha_1$ yields $d_{p1}=0$, $d_{p2}=3-4\alpha_{1}$ and $d_c+d_0=4\alpha_{1}$. If there is no common part split from the unicast messages, i.e., $d_c=0$, we have $d_{1}=0$, $d_{2}=3-4\alpha_{1}$ and $d_0=4\alpha_{1}$, which is actually $\mathcal{P}_{24}$.
To achieve corner point $\mathcal{P}_{14}'$, substituting $A_1=0$ and $A_2=\alpha_{1}$ into \eqref{Q1}-\eqref{Q4} yields $d_{p1}=0$, $d_{p2}=2\alpha_{1}$ and $d_c+d_0=2$. If there is no common part split from the unicast messages, i.e., $d_c=0$, we have $d_{1}=0$, $d_{2}=2\alpha_{1}$ and $d_0=2$, which is actually $\mathcal{P}_{14}'$.

\subsubsection{CASE-4 (If $ \alpha_{1} \leq 1-\alpha_2$)}
Via GENERAL CASE-4,  the DoF region of this $(4, 2, 3)$ MIMO BC with imperfect CSIT and hybrid message is simplified to	
\begin{eqnarray} \label{bound3}
\mathcal{D} =	\{	(d_1,d_2,d_0)  
\in \mathbb{R}_+^3| \qquad \qquad \qquad \qquad \qquad \quad \nonumber \\
\left.		 \begin{aligned}
& \ell_1,\ell_2,\ell_4 \text{ in } \eqref{bound1}, \\
& \ell_3: d_1+d_2+d_0 \le 3 + \dfrac{\alpha_1\alpha_2}{1+\alpha_2 - \alpha_1}.  
%\ell_5: \dfrac{d_1}{N_1} + \dfrac{d_2+d_0}{N_2} \leq 1 + \dfrac{M-N_1}{N_2}\alpha_1. \\	
\end{aligned}
\right\}. 
\end{eqnarray} 	

Note that the on-coordinate corner points are given by $(2,0,0)$, $(0,3,0)$, and $(0,0,2)$, which can be trivially achieved.
According to Proposition 4, all off-coordinate corner points are given as follows. $	\mathcal{P}_{234}=\frac{1}{1+\alpha_2 - \alpha_1}(\alpha_1\alpha_2, 4\alpha_1^2 - \alpha_1\alpha_2 -7\alpha_1 + 3\alpha_2 + 3, 4\alpha_1-4\alpha_1^2+4\alpha_1\alpha_2)$ if $\alpha_1 \le \frac{1}{2} + \frac{1}{2}\alpha_2$,
$\mathcal{P}_{123}= \frac{1}{1+\alpha_2 - \alpha_1}(\alpha_1\alpha_2, \alpha_2-\alpha_1+\alpha_1\alpha_2+1, 2\alpha_2 - 2\alpha_1 - \alpha_1\alpha_2 +2)$ if $\frac{\alpha_1\alpha_2}{1+\alpha_2 - \alpha_1} \le \min \{2\alpha_1 - 1, 2\}$,
$\mathcal{P}_{124} = (2\alpha_1 - 1, 2\alpha_1, 3-2\alpha_1)$ if $2\alpha_1 - 1 \le \frac{\alpha_1\alpha_2}{1+\alpha_2 - \alpha_1}$ and $\frac{1}{2} \le \alpha_1 \le \frac{3}{2}$.
In Fig. 6, we illustrate the DoF region with $\alpha_1=0.6$ and $\alpha_{2}=0.4$, where $\mathcal{P}_{124}$ and $\mathcal{P}_{234}$ exist. 
In addition, $\mathcal{P}_{14}=(2,2\alpha_{1},0),  
\mathcal{P}_{34}=(\frac{2\alpha_1(\alpha_2 - 2\alpha_1 +2)}{1+\alpha_2 - \alpha_1},\frac{3\alpha_{1}\alpha_{2}}{1+\alpha_2 - \alpha_1}-4\alpha_{1}+3,0),  
\mathcal{P}_{23}=(\frac{\alpha_{1}\alpha_{2}}{1+\alpha_2 - \alpha_1},3,0),  
\mathcal{P}_{14}'=(0,2\alpha_{1},2)$, and 
$\mathcal{P}_{24}=(0,3-4\alpha_{1},4\alpha_{1})$. 
We next show that the above off-coordinate corner points are achievable.

In this case, the space-time transmission scheme provided by \cite{hao2017achievable} is used.  
Suppose there are totally $T$ time slots during the transmission, and $T$ is sufficiently large. 
%	The power exponents in time slot $l$ is $(A_{1,l}, A_{2,l})$.
%\begin{equation}
%	(A_{1,l},A_{2,l})= 
%	\begin{cases} 
%		(\alpha_2,1),       & {l = 1,...,\rho T},\\
%		(\alpha_{2},\alpha_{1}),      & {l = \rho T + 1,...,T},\\
%	\end{cases}   
%\end{equation} 
%	where $\rho \in [0,1]$.  
By \eqref{PowerEx}, 
the optimal ratio $\rho^*$ to maximize the achievable sum-DoF is given by
$
\rho^*=\frac{1-2\alpha_{1}+\alpha_{2}}{1+\alpha_2 - \alpha_1}
$.
Therefore, this space-time transmission scheme can achieve
\begin{subequations}
\begin{eqnarray}
&&	d_{p1} \le \frac{\alpha_1\alpha_2}{1+\alpha_2 - \alpha_1}, \\
&&	d_{p2} \le  \frac{4\alpha_1^2 - \alpha_1\alpha_2 -7\alpha_1 + 3\alpha_2 + 3}{1+\alpha_2 - \alpha_1}, \\
&&	d_{c} + d_0 \le \frac{-3\alpha_1^2 + 4\alpha_1 + \alpha_1\alpha_2}{1+\alpha_2 - \alpha_1}.
\end{eqnarray}
\end{subequations}
To achieve corner points  $\mathcal{P}_{234}$, $\mathcal{P}_{123}$, $\mathcal{P}_{34}$ and $\mathcal{P}_{23}$, we apply the space-time transmission scheme. $(A_{1,l},A_{2,l})$ to achieve corner points $\mathcal{P}_{34}$ and $\mathcal{P}_{23}$ are given in \cite{hao2017achievable} with $d_0 = 0$. In particular, to achieve corner point $\mathcal{P}_{234}$, we consider there is no common part split from unicast messages, i.e., $d_c=0$, and $d_0 = \frac{4\alpha_1-4\alpha_1^2+4\alpha_1\alpha_2}{1+\alpha_2-\alpha_1}$, then corner point $\mathcal{P}_{234}$ is achieved.  To achieve corner point $\mathcal{P}_{123}$, we consider there is no common part split from unicast messages, i.e., $d_c=0$, $d_{p1} = \frac{\alpha_2-\alpha_1+\alpha_1\alpha_2+1}{1+\alpha_2-\alpha_1}$, $d_0 = \frac{2\alpha_2 - 2\alpha_1 - \alpha_1\alpha_2 +2}{1+\alpha_2-\alpha_1}$, then corner point $\mathcal{P}_{123}$ is achieved. 	The
corner points $\mathcal{P}_{14}$, $\mathcal{P}_{14}'$ and $\mathcal{P}_{24}$ are the same as CASE-3, thus with the same solutions to achieve. The corner point $\mathcal{P}_{124}$ is the same as CASE-2, thus the achievability can be guaranteed accordingly.

%\item 

%\end{itemize}

%It worth mentioning that corner points $\mathcal{P}_{234}$, $\mathcal{P}_{34}$ and $\mathcal{P}_{23}$ cannot be achieved without space-time transmission. If we apply $A_1=A_1^*=\alpha_{2}$ and $A_2=A_2^*=\frac{1+\alpha_{2}}{2}$, it yields $d_{p1}=\alpha_1+\frac{\alpha_2}{2}-\frac{1}{2}$, $d_{p2}=\frac{3\alpha_2}{2}-\alpha_1+\frac{3}{2}$,
%$d_c+d_0=\alpha_1-\frac{3\alpha_{2}}{2}+\frac{3}{2}$,
%then the achievable sum-DoF is $\frac{5+\alpha_{2}+2\alpha_{1}}{2}$, which is smaller than $3+\frac{\alpha_{1}\alpha_{2}}{1+\alpha_2 - \alpha_1}$ with space-time transmission \cite{hao2017achievable}.

%\subsubsection{CASE-5 (When $\alpha_{1} \leq \frac{1 }{2}$)}
%The DoF region in this case is the same as that in CASE-4, with the same achieving solution (corner points). Note that with a different CSIT quality, the DoF region without space-time transmission is different. It can be seen that the power levels maximize the achievable sum-DoF in \eqref{sumDoF} are given by
%\begin{equation}
%	\begin{cases}
%		A_{1}^{*}=\alpha_{2}, \\
%		A_{2}^{*}=1-\alpha_{1}.
%	\end{cases}	 
%\end{equation}
%Substituting $A_1=A_1^*=\alpha_{2}$ and $A_2=A_2^*=1-\alpha_{1}$ into \eqref{Q1}, \eqref{Q2}, \eqref{Q3}, and \eqref{Q4} yields $d_{p1}=0$, $d_{p2}=3-4\alpha_{1}$ and $d_c+d_0=4\alpha_{1}$. Therefore, in this case the sum-DoF without space-time transmission is 3.

\begin{figure}[h]
\centering
\includegraphics[width=6.5cm]{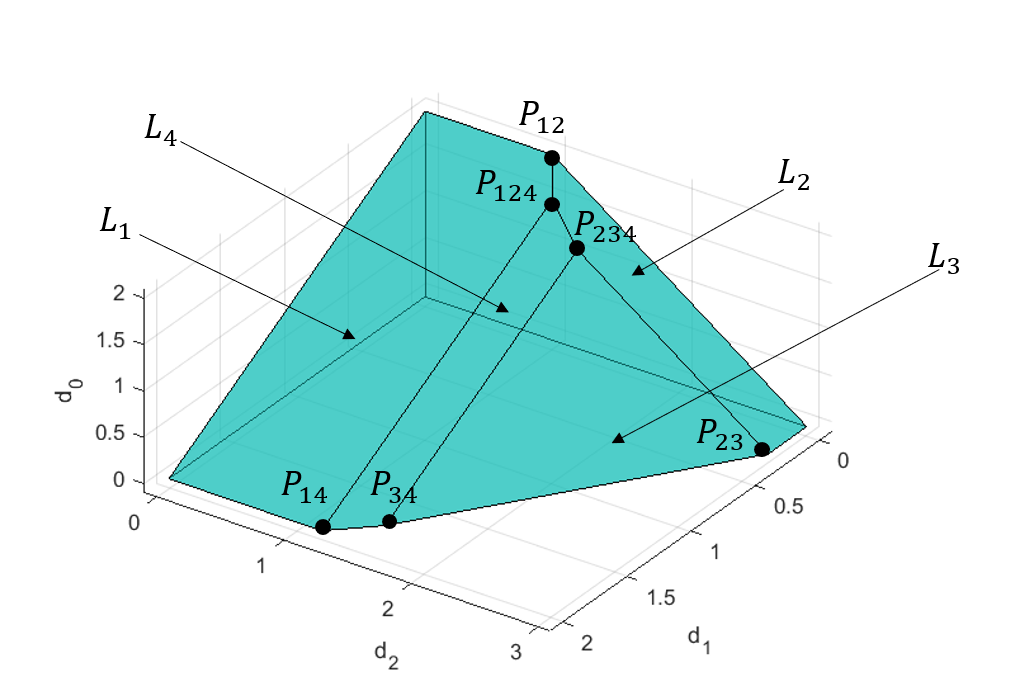}
\caption{The achievable DoF region of $(4,2,3)$ MIMO BC with $\alpha_1=0.6$ and $\alpha_{2}=0.4$.} \label{F6}
\end{figure}
\section{Converse Proof of Theorem 1}
The DoF outer region can be derived based on the following three steps: In the first step, let us relax the decoding requirement of $W_0$ and only request $\text{Rx}_1$ to decode it, such that $W_0$ degenerates into $W_1$. Since the relaxation won’t hurt the DoF region with hybrid messages, the DoF region of this new channel
is an outer bound of that of the original one.  Hence, according to \cite[Proposition 1]{hao2017achievable}, we obtain the following outer bound. Note that \cite[Remark 2]{davoodi2020degrees} states that the achievable DoF region in \cite[Proposition 1]{hao2017achievable} matches the DoF  outer region in \cite[Theorem 2]{hao2017achievable}, which shows the DoF-optimality.
\begin{eqnarray} \label{outer1}	
\mathcal{D}^\text{outer}_1=\left\{(d_1,d_2,d_0)\in\mathbb{R}_+^3| \right. \qquad \qquad \qquad \qquad \qquad \nonumber\\ 
\left. \begin{aligned}   
&d_1+d_0 \le \min\{M,N_1\}, \\
&d_2 \le \min\{M,N_2\}, \\
&d_1+d_2+d_0 \le \min\{M,N_2\} + \\
& \quad \quad [\min\{M,N_1+N_2\}-\min\{M,N_2\}]\alpha_0, \\
&	\dfrac{d_1+d_0}{\min\{M,N_1\}} + \dfrac{d_2}{\min\{M,N_2\}} \leq
1 + \\
& \quad \quad \dfrac{\min\{M,N_1+N_2\} - \min\{M,N_1\}} {\min\{M,N_2\}} \alpha_1, \\
&	\dfrac{d_1}{\min\{M,N_1\}} + \dfrac{d_2}{\min\{M,N_2\}}  \leq
1 + \\
& \quad \quad \dfrac{\min\{M,N_1+N_2\} - \min\{M,N_1\}} {\min\{M,N_2\}} \alpha_1. \\ 	
\end{aligned} \right\},    
\end{eqnarray}	

In the second step, we relax the decoding requirement of $W_0$ and only require $\text{Rx}_2$ to decode it, such that $W_0$ degenerates into $W_2$. Since the relaxation won't hurt the DoF region with hybrid messages, the DoF region of this new channel is an outer bound of that of the original one. Thus, according to \cite[Proposition 1]{hao2017achievable} we obtain the following outer bound. 
\begin{eqnarray} \label{outer2} 
\mathcal{D}^\text{outer}_2 = \left\{(d_1, d_2, d_0) \in \mathbb{R}_+^3|\right. \qquad \qquad \qquad \qquad \qquad \nonumber   \\	
\left. \begin{aligned}
&	d_1 \le \min\{M,N_1\}, \\
& 	d_2+d_0 \le \min\{M,N_2\}, \\
&	d_1+d_2+d_0 \le \min\{M,N_2\} + \\
& \quad \qquad [\min\{M,N_1+N_2\} - \min\{M,N_2\}] \alpha_0,  \\
& 	\dfrac{d_1}{\min\{M,N_1\}} + \dfrac{d_2}{\min\{M,N_2\}} \leq
1 + \\
& \qquad \qquad  \dfrac{\min\{M,N_1+N_2\} - \min\{M,N_1\}} {\min\{M,N_2\}} \alpha_1, \\
& 	\dfrac{d_1}{\min\{M,N_1\}} + \dfrac{d_2+d_0}{\min\{M,N_2\}}  \leq
1 + \\
& \qquad \qquad  \dfrac{\min\{M,N_1+N_2\} - \min\{M,N_1\}} {\min\{M,N_2\}} \alpha_1.  	
\end{aligned} \right\}.
\end{eqnarray}

Finally, after union of $ \mathcal{D}^\text{outer}_1$ and $\mathcal{D}^\text{outer}_2$, it can be seen that $	d_1 + d_0 \le \min\{M,N_1\}$ is dominated and replaced by $	d_1 \le \min\{M,N_1\}$, $d_2 + d_0 \le \min\{M,N_2\}$ is dominated and replaced by $d_2 \le \min\{M,N_2\}$, $\frac{d_1}{\min\{M,N_1\}} + \frac{d_2}{\min\{M,N_2\}} \leq
1 + \frac{\min\{M,N_1+N_2\} - \min\{M,N_1\}} {\min\{M,N_2\}} \alpha_1$ is dominated and replaced by $\frac{d_1+d_0}{\min\{M,N_1\}} + \frac{d_2}{\min\{M,N_2\}} \leq
1 + \frac{\min\{M,N_1+N_2\} - \min\{M,N_1\}} {\min\{M,N_2\}} \alpha_1$, $\frac{d_1}{\min\{M,N_1\}} + \frac{d_2}{\min\{M,N_2\}}  \leq
1 + \frac{\min\{M,N_1+N_2\} - \min\{M,N_1\}} {\min\{M,N_2\}} \alpha_1$ is dominated and replaced by $	\frac{d_1}{\min\{M,N_1\}} + \frac{d_2+d_0}{\min\{M,N_2\}}  \leq
1 + \frac{\min\{M,N_1+N_2\} - \min\{M,N_1\}} {\min\{M,N_2\}} \alpha_1$, $\frac{d_1}{\min\{M,N_1\}} + \frac{d_2+d_0}{\min\{M,N_2\}} \leq
1 + \frac{\min\{M,N_1+N_2\} - \min\{M,N_1\}} {\min\{M,N_2\}} \alpha_1$ is dominated and replaced by $\frac{d_1+d_0}{\min\{M,N_1\}} + \frac{d_2}{\min\{M,N_2\}} \leq
1 + \frac{\min\{M,N_1+N_2\} - \min\{M,N_1\}} {\min\{M,N_2\}} \alpha_1$. Therefore, the converse of DoF region can be derived. This ends the proof.

{	\section{The Proof of Theorem 2} 	
\subsection{Converse Proof} 
The converse proof of Theorem 2 is similar to the proof of Theorem 1, which exploits the results in \cite{zhang2022dof}. In the first step, we relax the decoding requirement of $W_0$ by only requesting $\text{Rx}_1$ to decode it and $W_0$ degenerates into $W_1$. An enhanced channel is therefore created. Based on the results of \cite{zhang2022dof}, a DoF outer region is given in the following. 
\begin{eqnarray}
\mathcal{Q}_1^\text{outer} = \{(d_1,d_2,d_0)  
\in \mathbb{R}_+^3 | \qquad \qquad \qquad \qquad \qquad \qquad \nonumber \\
\left.
\begin{aligned}
\frac{d_1+d_0}{\min\{N_1+\alpha_2N_2,M\}} 
+ \frac{d_2}{\min\{N_2,M\}} \le 1, \\
\frac{d_1+d_0}{\min\{N_1,M\}} + \frac{d_2}{\min\{N_2+\alpha_1N_1,M\}}   \le 1.
\end{aligned}
\right\}. \label{D-Outer-1}
\end{eqnarray}

In the second step, we relax the decoding requirement of $W_0$ by only requesting $\text{Rx}_2$ to decode it and $W_0$ degenerates into $W_2$. An enhanced channel is therefore created. Based on the results of \cite{zhang2022dof}, a DoF outer region is given in the following. 
\begin{eqnarray}
\mathcal{Q}_2^\text{outer} = \{(d_1,d_2,d_0)  
\in \mathbb{R}_+^3 | \qquad \qquad \qquad \qquad \qquad \qquad \nonumber \\
\left.
\begin{aligned}
\frac{d_1}{\min\{N_1+\alpha_2N_2,M\}} 
+ \frac{d_2+d_0}{\min\{N_2,M\}} \le 1, \\
\frac{d_1}{\min\{N_1,M\}} + \frac{d_2+d_0}{\min\{N_2+\alpha_1N_1,M\}}   \le 1.
\end{aligned}
\right\}. \label{D-Outer-2}
\end{eqnarray}

In the final step, we combine $\mathcal{Q}_1^\text{outer}$ and $	\mathcal{Q}_2^\text{outer}$ together, and observe that  $\frac{d_1+d_0}{\min\{N_1+\alpha_2N_2,M\}} 
+ \frac{d_2}{\min\{N_2,M\}} \le 1$ is dominated by $	\frac{d_1}{\min\{N_1+\alpha_2N_2,M\}} 
+ \frac{d_2+d_0}{\min\{N_2,M\}} \le 1$ and $\frac{d_1}{\min\{N_1,M\}} + \frac{d_2+d_0}{\min\{N_2+\alpha_1N_1,M\}}   \le 1$ is dominated by $	\frac{d_1+d_0}{\min\{N_1,M\}} + \frac{d_2}{\min\{N_2+\alpha_1N_1,M\}}   \le 1$.  After wiping out the above redundancy, we arrive at the desired result. 

\subsection{Achievability Proof}
In order to prove the achievability, it suffices to derive all corner points and provide their achievable schemes. First of all, there are three off-coordinate corner points, where one coordinate is set to zero. They are given by
$
\mathcal{P}_{12} =  (\frac{\min\{N_1+\alpha_{2}N_2,M\}\min\{N_1,M\}(\min\{N_2+\alpha_{1}N_1,M\}-\min\{N_2,M\})}{\min\{N_1+\alpha_{2}N_2,M\}\min\{N_2+\alpha_{1}N_1,M\}-\min\{N_2,M\}\min\{N_1,M\}}$,   $\frac{\min\{N_2+\alpha_{1}N_1,M\}\min\{N_2,M\}(\min\{N_1+\alpha_{2}N_2,M\}-\min\{N_1,M\})}{\min\{N_1+\alpha_{2}N_2,M\}\min\{N_2+\alpha_{1}N_1,M\}-\min\{N_2,M\}\min\{N_1,M\}},  0)
$. 	
If $\min\{N_2,M\} \le  \min\{\min\{N_1 + \alpha_2 N_2, M\}, \min\{N_1,M\}\}$, then $\mathcal{P}_{10} = \frac{1}{\min\{N_2,M\} - \min\{N_1+\alpha_2 N_2,M\}}((\min\{N_2,M\}-\min\{N_1,M\})\min\{N_1+\alpha_2 N_2, M\}, 0,$$ (\min\{N_1,M\}-\min\{N_1+\alpha_2 N_2, M\})\min\{N_2,M\})$ exists. Furthermore, if $\min\{N_1,M\} \le  \min\{\min\{N_2 + \alpha_1 N_1, M\}, \min\{N_2,M\}\}$, then $\mathcal{P}_{20} = \frac{1}{\min\{N_1,M\} - \min\{N_2+\alpha_1 N_1,M\}}(0,(\min\{N_1,M\}-\min\{N_2,M\})\min\{N_2+\alpha_1 N_1, M\} ,(\min\{N_2,M\}-\min\{N_2+\alpha_1 N_1, M\})\min\{N_1,M\})$ exists.  

The corner point $\mathcal{P}_{12}$ is the intersection of the following three hyperplanes.
\begin{equation}
\begin{cases}
\dfrac{d_1}{\min\{N_1+\alpha_2N_2,M\}} + \dfrac{d_2+d_0}{\min\{N_2,M\}} = 1, \\
\dfrac{d_1+d_0}{\min\{N_1,M\}} + \dfrac{d_2}{\min\{N_2+\alpha_1N_1,M\}} = 1, \\
d_0=0. 
\end{cases}		
\end{equation}
Next, we show that $\mathcal{P}_{12}$ is a valid corner point by the following reasons. Since
$\min\{N_1+\alpha_2N_2,M\}\geq\min\{N_1,M\}$ and 
$\min\{N_2+\alpha_1N_1,M\}\geq\min\{N_2,M\}$,
we have 
$
\min\{N_2+\alpha_{1}N_1,M\}-\min\{N_2,M\} \geq 0
$, 
$
\min\{N_1+\alpha_{2}N_2,M\}\min\{N_2+\alpha_{1}N_1,M\}-\min\{N_2,M\}\min\{N_1,M\} \geq 0
$, 
$
\min\{N_1+\alpha_{2}N_2,M\}-\min\{N_1,M\} \geq 0
$, and 
$
\min\{N_1+\alpha_{2}N_2,M\}\min\{N_2+\alpha_{1}N_1,M\}-\min\{N_2,M\}\min\{N_1,M\} \geq 0
$.
Therefore, all the coordinates of $\mathcal{P}_{12}$ are non-negative, indicating it a valid corner point. The achievable scheme for $\mathcal{P}_{12}$ is the same as that in \cite{zhang2022dof} without multicast message.

The corner point $\mathcal{P}_{10}$ is the intersection of the following three hyperplanes.
\begin{equation}
\begin{cases}
\dfrac{d_1}{\min\{N_1+\alpha_2N_2,M\}} + \dfrac{d_2+d_0}{\min\{N_2,M\}} = 1, \\
\dfrac{d_1+d_0}{\min\{N_1,M\}} + \dfrac{d_2}{\min\{N_2+\alpha_1N_1,M\}} = 1, \\
d_2=0. 
\end{cases}		
\end{equation}
Next, we show that $\mathcal{P}_{10}$ is a valid corner point by the following reasons. Since $N_1\geq N_2$, it can be seen that $\min\{N_1,M\}\geq\min\{N_2,M\}$, $\min\{N_1+\alpha_1N_2,M\}\geq\min\{N_2,M\}$, and $\min\{N_1+\alpha_2N_2,M\}\geq\min\{N_1,M\}$. Thus, all the coordinates of $\mathcal{P}_{10}$ are non-negative, indicating it a valid corner point. The achievable scheme for $\mathcal{P}_{10}$ is sending $\frac{(\min\{N_2,M\}-\min\{N_1,M\})\min\{N_1+\alpha_2 N_2, M\}}{\min\{N_2,M\} - \min\{N_1+\alpha_2 N_2,M\}}$ DoF for $W_1$ and simultaneously $\frac{ (\min\{N_1,M\}-\min\{N_1+\alpha_2 N_2, M\})\min\{N_2,M\}}{\min\{N_2,M\} - \min\{N_1+\alpha_2 N_2,M\}}$ DoF for $W_0$ by broadcasting. 

The corner point $\mathcal{P}_{20}$ is the intersection of the following three hyperplanes.
\begin{equation}
\begin{cases}
\dfrac{d_1}{\min\{N_1+\alpha_2N_2,M\}} + \dfrac{d_2+d_0}{\min\{N_2,M\}} = 1, \\
\dfrac{d_1+d_0}{\min\{N_1,M\}} + \dfrac{d_2}{\min\{N_2+\alpha_1N_1,M\}} = 1, \\
d_1=0. 
\end{cases}		
\end{equation}
Next, we show that $\mathcal{P}_{20}$ is a valid corner point by the following reasons. Since $N_2\geq N_1$, it can be seen that $\min\{N_2,M\}\geq\min\{N_1,M\}$, $\min\{N_2+\alpha_1N_1,M\}\geq\min\{N_1,M\}$, and $\min\{N_2+\alpha_1N_1,M\}\geq\min\{N_2,M\}$. Thus, all the coordinates of $\mathcal{P}_{20}$ are non-negative, indicating it a valid corner point.  The achievable scheme for $\mathcal{P}_{20}$ is sending $\frac{(\min\{N_1,M\}-\min\{N_2,M\})\min\{N_2+\alpha_1 N_1, M\}}{\min\{N_1,M\} - \min\{N_2+\alpha_1 N_1,M\}}$ DoF for $W_2$ and simultaneously $\frac{(\min\{N_2,M\}-\min\{N_2+\alpha_1 N_1, M\})\min\{N_1,M\}}{\min\{N_1,M\} - \min\{N_2+\alpha_1 N_1,M\}}$ DoF for $W_0$ by broadcasting.

For other corner points on the coordinate, i.e., $\mathcal{P}_1 = \left(\min\{M,N_1\},0,0\right)$, $\mathcal{P}_2 = \left(0,\min\{M,N_2\},0\right)$, and 
$\mathcal{P}_0 = \left(0,0,\min\{N_1,N_2\}\right)$, they can be achieved by forcing two of $(d_1, d_2, d_0)$ in (\ref{Delayed}) to zero while solving the remaining. For the validation, it is quite intuitive that $\min\{M,N_1\}\geq0$, $\min\{M,N_2\}\geq0$, and $\min\{N_1,N_2\}\geq0$, so that all the coordinates of these points are non-negative. In addition, since 
$\min\{M,N_1\}\leq\min\{M,N_1+\alpha_2N_2\}$, 
$\min\{M,N_2\}\leq\min\{M,N_2+\alpha_1N_1\}$, 
$\min\{N_1,N_2\}\leq\min\{N_1,M\}$, and 
$\min\{N_1,N_2\}\leq\min\{N_2,M\}$, neither
$\mathcal{P}_1$, $\mathcal{P}_2$ nor $\mathcal{P}_0$ violates (\ref{Delayed}). Thus, $\mathcal{P}_1$, $\mathcal{P}_2$ and $\mathcal{P}_0$ are valid corner points. The achievable schemes for $\mathcal{P}_1$, $\mathcal{P}_2$ and $\mathcal{P}_0$ are simply broadcasting the corresponding messages for each time slot.
}

{	\section{Extension to $K$-user Setups}
To extend our results to more general $K$-user MIMO BC setups, we 
can leverage the converse technique. That is, supposing that the DoF region of $K$-user MIMO BC with imperfect CSIT and private messages is obtained, the target DoF region with hybrid messages can be derived in the following, which is also equivalent to establishing the DoF outer region. First, the derived DoF region of $K$-user MIMO BC with imperfect CSIT and private messages is given by
\begin{eqnarray}
&& \!\!\!\!\!\!\!\!\!\!\!\!\!\!\!\!\!\!\!\!\!\! \mathcal{D}^K :=   \left\{(d_1,
\cdots,d_K) \in \mathbb{R}_+^K \left| \begin{split}\ell_1(d_1,\cdots,d_K) \le 0,   \\
\vdots \qquad \qquad   \\
\ell_L(d_1,\cdots,d_K) \le 0. \end{split}\right.\right\},
\end{eqnarray}
where there are $L$ inequalities constituting the region. Second, supposing receiver $1$ needs to decode the additional multicast message, we have 
\begin{eqnarray}
&&	\!\!\!\!\!\!\!\!\!\!\!\!\!\!\!\!\!\!\!\!\!\!	\mathcal{D}_1^{K+1} :=   \left\{\begin{split}(d_1,
\cdots,d_K, \\ d_0)\in \mathbb{R}_+^{K+1} \end{split} \left| \begin{split}\ell_1(d_1+d_0,\cdots,d_K) \le 0,   \\
\vdots \qquad \qquad   \\
\ell_L(d_1+d_0,\cdots,d_K) \le 0. \end{split}\right.\right\}.  
\end{eqnarray}
It can be seen that the target DoF region with hybrid messages is contained in $\mathcal{D}_1^{K+1}$, since the decodability constraint is relaxed. Likewise, we denote $\mathcal{D}_k^{K+1},k=1,\cdots,K$ as the DoF region that receiver $k$ needs to decode the additional multicast message. Finally, the 
target DoF region of $K$-user MIMO BC with imperfect CSIT and hybrid messages is given by 
\begin{equation}
\mathcal{D}^{K+1} := \mathcal{D}_1^{K+1} \cap \mathcal{D}_2^{K+1} \cap \cdots \cap \mathcal{D}_K^{K+1}, 
\end{equation}
which is the union of all relaxations. A simplified expression of $\mathcal{D}^{K+1}$ occurs after removing all redundancy. Nevertheless, to prove the achievability of this region, it suffices to figure out all corner points and show their achievable schemes, which is highly non-trivial, due to curse of dimensionality.
}

\section{Conclusion}
In this paper, we characterized the DoF region of the  two-user  MIMO BC with hybrid messages and imperfect CSIT, which was given by the tight converse and achievability proof of the DoF region. Specifically, the proposed hybrid message-aware rate-splitting scheme achieves the DoF converse. Besides, we obtained the sum-DoF of the two-user MIMO BC with hybrid messages and imperfect CSIT.  We showed that the DoF region with hybrid messages is with specific three-dimensional structure w.r.t. antenna configurations and CSIT qualities. To achieve the strictly positive corner points, we further showed that it is unnecessary to split the unicast messages into private and common messages, since the allocated power for the common part should be zero. We also derived the DoF region of the two-user MIMO BC with hybrid messages and delayed CSIT in the presence of errors. 

Aside from information theoretic significance stated above, the communication theoretic significance inspired by DoF analysis are summarized as follows. 1) The achievable sum-rate of two-user MIMO BC with hybrid messages and imperfect CSIT will not be higher than the same system with unicast messages only, when SNR is asymptotically large. 2) Due to the transmission of a multicast message, the achievable rate of unicast messages deteriorates, compared with the same system with unicast messages only.  

There are many extensions worthwhile to be investigated. For instance, we would study the rate-splitting of $K$-user MIMO BC with imperfect CSIT and  hybrid messages, and two-user MIMO interference channel with imperfect CSIT and hybrid messages. Furthermore, the rate optimization with finite SNR is also an interesting problem.

\begin{appendices}

\section{Proof of Propositions 1, 2, and 4}

\subsection{Proof of Proposition 1}

First of all, we figure out all corner points with one zero coordinate.
According to \cite{hao2017achievable}, corner points without multicast messages (i.e., $d_0 = 0$) are given by  $\mathcal{P}_{13} = (N_1,N_2-N_1+(M-N_2)\alpha_{2},0)$  and $\mathcal{P}_{23} = ((M-N_2)\alpha_{2},N_2,0)$. If $d_1 = 0$, it turns out that $\mathcal{D} = \{(d_2,d_0)\in \mathbb{R}_+^2|d_0 \le N_1,d_2+d_0\le N_2, \frac{d_2}{N_2}+ \frac{d_0}{N_1} \le 1 + \frac{M-N_1}{N_2}\alpha_1 \}$, where the off-coordinate corner point is given by $\mathcal{P}_{12} = (0,N_2-N_1,N_1)$. If $d_2 = 0$, it turns out that $\mathcal{D} = \{(d_1,d_0)\in \mathbb{R}_+^2|d_0 \le N_2, d_1+d_0 \le N_1\}$, where the off-coordinate corner point does not exist due to $N_1 - N_2 \le 0$.

Next, we derive the corner points with strictly positive coordinate. However, although there are $\mathcal{P}_{123}$, $\mathcal{P}_{124}$, $\mathcal{P}_{234}$, and $\mathcal{P}_{134}$, only $\mathcal{P}_{123}$ is qualified as a corner point. The reason is shown in the following. Via \textit{MATLAB Symbolic Calculation}, corner point with intersection of $\ell_1$, $\ell_2$, and $\ell_3$ is given by 
$
\mathcal{P}_{123} = ((M-N_2)\alpha_{2},N_2-N_1+(M-N_2)\alpha_{2},N_1-(M-N_2)\alpha_{2}).
$ It can be check that $\mathcal{P}_{123}$ satisfies $\ell_4$, since
$
\frac{d_1+d_0}{N_1} + \frac{d_2}{N_2} = 1 + \frac{N_2- N_1+(M-N_2)\alpha_{2}}{N_2}  \nonumber 
$,
then 
\begin{eqnarray} 
&&	\!\!\!\!\!\!\!\!\!\! \frac{d_1+d_0}{N_1} + \frac{d_2}{N_2} - \left(1 + \frac{(M-N_1)\alpha_1}{N_2} \right) \nonumber \\
&& \!\!\!\!\!\!\!\!\!\! = \frac{N_2- N_1+(M-N_2)\alpha_{2} - (M-N_1)\alpha_1}{N_2}   \overset{(a)}{\le} 0,	 
\end{eqnarray} where (a) is due to $\alpha_{1} \geq \frac{N_2-N_1+(M-N_2)\alpha_{2}}{M-N_1}$.
The corner point with intersection of $\ell_1$, $\ell_2$, and $\ell_4$ is given by
$
\mathcal{P}_{124} = (N_1-N_2+(M-N_1)\alpha_1, (M-N_1)\alpha_1, N_2-(M-N_1)\alpha_1).
$
It can be checked that $\mathcal{P}_{124}$ violates $\ell_3$ and does not exist, since
$
d_1 + d_2 + d_0  = N_1-N_2+(M-N_1)\alpha_1+ (M-N_1)\alpha_1+ N_2-(M-N_1)\alpha_1   \nonumber
$,
then 
\begin{eqnarray} 
&&	\!\!\!\!\!\!\!\!\!\! d_1 + d_2 + d_0 - (N_2 + (M-N_2)\alpha_2)  \nonumber \\
&& \!\!\!\!\!\!\!\!\!\! = N_1 + (M-N_1)\alpha_1 - N_2 - (M-N_2)\alpha_2  \overset{(a)}{\ge}0,   
\end{eqnarray}
where (a) is due to $\alpha_{1} \geq \frac{N_2-N_1+(M-N_2)\alpha_{2}}{M-N_1}$. 
corner point with intersection of  $\ell_2$, $\ell_3$, and $\ell_4$ is given by
$
\mathcal{P}_{234} = ((M-N_2)\alpha_2, \frac{N_2(M-N_2)\alpha_2-N_1(M-N_1)\alpha_1+N_2(N_2-N_1)}{N_2 -N_1},  \\
\frac{N_1(M-N_1)\alpha_1-N_2(M-N_2)\alpha_2}{N_2-N_1}).
$
It can be checked that $\mathcal{P}_{234}$ violates $\ell_1$ and does not exist, since
$		d_1+d_0  =   \frac{N_1(M-N_1)\alpha_1-N_2(M-N_2)\alpha_2}{N_2-N_1}  
$ $+ (M-N_2)\alpha_2$,
then
\begin{eqnarray} 
&& \!\!\!\!\!\!\!\!\!\! d_1+d_0 - N_1 \nonumber \\
&& \!\!\!\!\!\!\!\!\!\! =   N_1 \frac{(M-N_1)\alpha_1 - (M-N_2)\alpha_2}{N_2 - N_1} - N_1 \overset{(a)}{\ge} 0, 
\end{eqnarray}
where (a) is due to $\alpha_{1} \geq \frac{N_2-N_1+(M-N_2)\alpha_{2}}{M-N_1}$. The corner point with intersection $\ell_1, \ell_3$, and $\ell_4$ does not exist, since the determinant of coefficient matrix for related equations is zero. %i.e.,
%\begin{equation}
%	\left| 
%	\begin{matrix}
%		1 & 0 & 1 \\
%		1 & 1 & 1 \\
%		1/N_1 & 1/N_2 & 1/N_1  
%	\end{matrix}
%	\right| = 0. \nonumber 
%\end{equation}
%This completes the proof.

\subsection{Proof of Proposition 2}
First of all, we figure out all corner points with one zero coordinate.
According to \cite{hao2017achievable}, corner points without multicast messages (i.e., $d_0 = 0$) are given by 
$\mathcal{P}_{14}=(N_1,(M-N_1)\alpha_{1},0)$, $\mathcal{P}_{34}=(N_1\alpha_1-\frac{N_1(M-N_2)}{M-N_1}\alpha_{2}+\frac{N_1(M-N_2)}{M-N_1},   -(N_1+N_2-M)\alpha_{1} + \frac{N_2(M-N_2)}{M-N_1}\alpha_{2}   +\frac{N_2(N_2-N_1)}{M-N_1},0)$, and
$\mathcal{P}_{23}=((M-N_2)\alpha_{1}+\frac{(M-N_2)(N_2-N_1)}{M-N_1}\alpha_{2}-\frac{(M-N_2)(N_2-N_1)}{M-N_1} ,N_2,0)$. If $d_1 = 0$, it turns out that 
$\mathcal{D} = \{(d_2,d_0)\in \mathbb{R}_+^2|
d_0 \le N_1, 
d_2 + d_0 \le N_2, 
\frac{d_0}{N_1} + \frac{d_2}{N_2} \leq 1+\frac{M-N_1}{N_2}\alpha_1
\}$,
where the off-coordinate corner point is given by $\mathcal{P}_{12}=(0,N_2-N_1,N_1)$ due to $\frac{N_2-N_1}{M-N_1} \leq \alpha_{1}$. If $d_2=0$, it turns out that 
$\mathcal{D} = \{(d_1,d_0)\in \mathbb{R}_+^2|
d_1 + d_0 \le N_1, 
d_0 \le N_2
\}$,
where the off-coordinate corner point does not exist due to $N_1 - N_2 \le 0$.

Next, we derive corner points with strictly positive coordinate. However, although there are $\mathcal{P}_{123}$, $\mathcal{P}_{124}$, $\mathcal{P}_{234}$, and $\mathcal{P}_{134}$, only $\mathcal{P}_{124}$ and $\mathcal{P}_{234}$ are qualified as corner points. The reason is shown in the following. The corner point with intersection of $\ell_1$, $\ell_2$, and $\ell_4$ is given by $\mathcal{P}_{124}=((M-N_1)\alpha_{1}-(N_2-N_1),(M-N_1)\alpha_1,N_2-(M-N_1)\alpha_{1})$, and corner point with intersection of $\ell_2$, $\ell_3$, and $\ell_4$ is given by $\mathcal{P}_{234}= 
((M-N_2)\alpha_{1}+ \frac{(M-N_2)(N_2-N_1)}{M-N_1}\alpha_{2}-\frac{(M-N_2)(N_2-N_1)}{M-N_1},   -(N_1+N_2-M)\alpha_{1}+    \frac{N_2(M-N_2)}{M-N_1}\alpha_{2}  +\frac{N_2(N_2-N_1)}{M-N_1}, (N_1+N_2-M)\alpha_{1}-\frac{(M-N_2)N_2}{M-N_1}\alpha_{2}+\frac{N_2(M-N_2)}{M-N_1})$. 

\begin{figure*}
\begin{eqnarray}
&& d_1 + d_2 + d_0 -\left(N_2+(M-N_2)\left(-\frac{N_2-N_1}{M-N_1}+\frac{N_2-N_1}{M-N_1}\alpha_{2}+\alpha_{1}\right) \right) \nonumber
\\
&& =(M-N_1)\alpha_{1} + N_1 - \left(N_2+(M-N_2)\left(-\frac{N_2-N_1}{M-N_1}+\frac{N_2-N_1}{M-N_1}\alpha_{2}+\alpha_{1}\right)\right)\nonumber \\
&& =\frac{(N_2-N_1)(N_2-N_1+(N_1-M)\alpha_{1}+(M-N_2)\alpha_{2})}{M-N_1}
\nonumber \\
&& \overset{(a)}{\le} \frac{(N_2-N_1)(N_2-N_1+(N_1-M)(\frac{N_2-N_1+(M-N_2)\alpha_{2}}{M-N_1})+(M-N_2)\alpha_{2})}{M-N_1}
\nonumber \\
&& = 
\frac{(N_2-N_1)(N_2-N_1- N_2+N_1-(M-N_2)\alpha_{2} +(M-N_2)\alpha_{2})}{M-N_1}  =0,
\label{pro2p124}
\end{eqnarray}
\hrule
\end{figure*}

It can be checked that $\mathcal{P}_{124}$ satisfies $\ell_3$, since 
$ 
d_1 + d_2 + d_0 = (M-N_1)\alpha_{1} + N_1 
$
, then we have (\ref{pro2p124}), where (a) is due to $\alpha_{1} \leq \frac{N_2-N_1+(M-N_2)\alpha_{2}}{M-N_1}$. 
It can also be checked that $\mathcal{P}_{234}$ satisfies $\ell_1$, since  
$
d_1+d_0 = N_1\alpha_{1} + \frac{(N_2-N_1)(2N_2-M)}{M-N_1} 
$
, then we have (\ref{pro2p234}), where (a) is due to $\alpha_{1} \leq \frac{N_2-N_1+(M-N_2)\alpha_{2}}{M-N_1}$. 
\begin{figure*}
\begin{eqnarray}
&& d_1+d_0 - N_1 \nonumber \\
&& =\frac{N_1(M - N_2 + (M-N_1)\alpha_{1} - (M-N_2)\alpha_{2})}{M - N_1} - N_1 \nonumber \\
&& \overset{(a)}{\le}   \frac{N_1(M - N_2 + (M-N_1)(\frac{N_2-N_1+(M-N_2)\alpha_{2}}{M-N_1}) - (M-N_2)\alpha_{2})}{M - N_1} - N_1 \nonumber \\
&& =   \frac{N_1(M - N_2 +  N_2-N_1+(M-N_2)\alpha_{2} - (M-N_2)\alpha_{2})}{M - N_1} -N_1
=0,\label{pro2p234}
\end{eqnarray} \hrule
\end{figure*}

\begin{figure*}
\begin{eqnarray}
&& \frac{d_1+d_0}{N_1}+\frac{d_2}{N_2} - \left(1+\frac{M-N_1}{N_2}\alpha_{1}\right) \nonumber  \\
&& =\frac{N_1^2+N_2^2-3N_1N_2+(2N_2-N_1)(N_1-M)\alpha_{1}+(N_2-N_1)(M-N_2)\alpha_{2}}{N_2(M-N_1)} \nonumber \\
&& \overset{(a)}{\ge}\frac{N_1^2+N_2^2-3N_1N_2+(2N_2-N_1)(N_1-M)(\frac{N_2-N_1+(M-N_2)\alpha_{2}}{M-N_1})+(N_2-N_1)(M-N_2)\alpha_{2}}{N_2(M-N_1)} \nonumber \\
&& =\frac{(M-N_2)(1-\alpha_{2})}{M-N_1} \overset{(b)}{\ge}0,
\label{pro2p123}
\end{eqnarray} \hrule
\end{figure*}
The corner point with intersection of $\ell_1$, $\ell_2$, and $\ell_3$ is given by $\mathcal{P}_{123}= (
(M-N_2)(\alpha_{1}+\frac{N_1-N_2}{M-N_1}-\frac{N_1-N_2}{M-N_1}\alpha_{2}),
N_2-N_1 + (M - N_2)(\alpha_{1} + \frac{N_1 - N_2}{M - N_1} - \frac{N_1-N_2}{M-N_1}\alpha_{2}),
N_1 - (M - N_2)(\alpha_{1} + \frac{N_1-N_2}{M-N_1} - \frac{N_1-N_2}{M-N_1}\alpha_{2})
)$.
$\mathcal{P}_{123}$ violates $\ell_4$ and does not exist, since
$\frac{d_1+d_0}{N_1}+\frac{d_2}{N_2} = 1 + \frac{N_2-N_1 + (M - N_2)(\alpha_{1} + \frac{N_1 - N_2}{M - N_1} - \frac{N_1-N_2}{M-N_1}\alpha_{2})}{N_2}$ then we have (\ref{pro2p123}), where (a) is due to $\alpha_{1} \leq \frac{N_2-N_1+(M-N_2)\alpha_{2}}{M-N_1}$, and (b) is due to $M \ge N_2, 1 \ge \alpha_{2},$ and $M \ge N_1$. The corner point with intersection $\ell_1, \ell_3$, and $\ell_4$ does not exist, since the determinant of coefficient matrix for related equations is zero. 

%This completes the proof.

%\subsection{Proof of Proposition 3}

\subsection{Proof of Proposition 4}
First of all, we figure out all corner points with one zero coordinate.
According to \cite{hao2017achievable}, corner points without multicast messages (i.e., $d_0 = 0$) are given by $\mathcal{P}_{14} = (N_1, (M-N_1)\alpha_1, 0)$, 	$\mathcal{P}_{34}=\frac{1}{\Delta}(N_1\alpha_1(M-N_1-(M-N_1)\alpha_1 + (M-N_2)\alpha_2)
,$ $
\left(M-N_{1}\right)N_1\alpha_1^2+( (M-N_2)(M-N_1-N_2)\alpha_2 + N_1^2 - N_2^2 - MN_1 + N_1N_2)\alpha_1 + (M-N_2)N_2\alpha_2 + N_2^2 - N_1N_2,  0)$, and  
$\mathcal{P}_{23}=(\frac{(M-N_2)^2\alpha_1\alpha_2}{\Delta}, N_2, 0),$ where $\Delta = (M-N_2)\alpha_2 + (N_2 - N_1)(1-\alpha_1)$. If $d_1 = 0$, it turns out that 
$\mathcal{D} = \{(d_2,d_0)\in \mathbb{R}_+^2|
d_0 \le N_1, 
d_2 + d_0 \le N_2, 
\frac{d_0}{N_1} + \frac{d_2}{N_2} \leq 1+\frac{M-N_1}{N_2}\alpha_1
\}$,
where the off-coordinate corner points are given by $\mathcal{P}_{14}'= (0,(M-N_1)\alpha_{1},N_1)$ and
$\mathcal{P}_{24}=(0,N_2-\frac{(M-N_1)N_1}{N_2-N_1}\alpha_{1},\frac{(M-N_1)N_1}{N_2-N_1}\alpha_{1})$. If $d_2=0$, it turns out that 
$\mathcal{D} = \{(d_1,d_0)\in \mathbb{R}_+^2|
d_1 + d_0 \le N_1, 
d_0 \le N_2
\}$, where the off-coordinate corner point does not exist due to $N_1 - N_2 \le 0$.

Next, we derive corner points with strictly positive coordinate. However, although there are $\mathcal{P}_{123}$, $\mathcal{P}_{124}$, $\mathcal{P}_{234}$, and $\mathcal{P}_{134}$,  while $\mathcal{P}_{234}$ is qualified as a corner point if $\alpha_1 \le \frac{N_2-N_1}{M-N_1} + \frac{M-N_2}{M-N_1}\alpha_2$,  $\mathcal{P}_{123}$ is qualified as a corner point if $\frac{(M-N_2)^2\alpha_1\alpha_2}\Delta \le \min\{(M-N_1)\alpha_1 - (N_2-N_1),N_1\}$, and $\mathcal{P}_{124}$ is qualified as a corner point if $(M-N_1)\alpha_1 - (N_2-N_1) \le \frac{(M-N_2)^2\alpha_1\alpha_2}\Delta$ and $\alpha_1 \le \frac{N_2}{M-N_1}$. The reason is shown in the following. The
corner point with intersection of $\ell_2$, $\ell_3$, and $\ell_4$ is given by $
\mathcal{P}_{234}=\frac{1}{\Delta}((M-N_2)^2\alpha_1\alpha_2,$ $
\left[M-N_{1}\right)N_1\alpha_1^2+( (M-N_2)(M-N_1-N_2)\alpha_2 + N_1^2 - N_2^2 - MN_1 + N_1N_2)\alpha_1 + (M-N_2)N_2\alpha_2 + N_2^2 - N_1N_2$, 	$((M-N_2)(N_1+N_2-M)\alpha_2 + (M-N_1)N_1(1-\alpha_1))\alpha_1)$. If $\alpha_1 \le \frac{N_2-N_1}{M-N_1} + \frac{M-N_2}{M-N_1}\alpha_2$, it can be checked that $\mathcal{P}_{234}$ satisfies $\ell_1$, since 
$
d_1+d_0 =  \frac{(M-N_2)N_1\alpha_1\alpha_2 + (M-N_1)N_1(1-\alpha_1)\alpha_1}\Delta 
$,
then we have (\ref{pro4p234}), where $(a)$ is due to $\alpha_1 \le \frac{N_2-N_1}{M-N_1} + \frac{M-N_2}{M-N_1}\alpha_2$.
\begin{figure*}
\begin{eqnarray}
&& d_1+d_0 - N_1 \nonumber  \\
&& = \frac{  (M-N_1)N_1(1-\alpha_1)\alpha_1 - (M-N_2)N_1\alpha_2(1 - \alpha_1) - (N_2 - N_1)N_1(1-\alpha_1)}\Delta \nonumber  \\
&& = N_1(1-\alpha_1)\frac{(M-N_1)\alpha_1 - (M-N_2)\alpha_2 - (N_2-N_1) }\Delta   \overset{(a)}{\le} 0,  
\label{pro4p234}
\end{eqnarray}
\hrule
\end{figure*}
Otherwise, if $  \frac{N_2-N_1}{M-N_1} + \frac{M-N_2}{M-N_1}\alpha_2 < \alpha_1$, it can be checked by the same steps that $\mathcal{P}_{234}$ violates $\ell_1$ and does not exist. The corner point with intersection of $\ell_1,\ell_2$, and $\ell_3$ is given by $\mathcal{P}_{123} = (\frac{(M-N_2)^2\alpha_1\alpha_2}\Delta,N_2-N_1 + \frac{(M-N_2)^2\alpha_1\alpha_2}\Delta, N_1 - \frac{(M-N_2)^2\alpha_1\alpha_2}\Delta)$, which requires $\frac{(M-N_2)^2\alpha_1\alpha_2}\Delta \le N_1$ for non-negativity of $d_0$. If $  \frac{(M-N_2)^2\alpha_1\alpha_2}\Delta \le (M-N_1)\alpha_1 - (N_2-N_1)$, $\mathcal{P}_{123}$ satisfies $\ell_4$, since $
\frac{d_1 + d_0}{N_1} + \frac{d_2}{N_2} = 1 + \frac{N_2-N_1 + \frac{(M-N_2)^2\alpha_1\alpha_2}\Delta}{N_2}$,
then 
\begin{eqnarray}
&& \frac{d_1 + d_0}{N_1} + \frac{d_2}{N_2} - \left( 1 + \frac{M-N_1}{N_2}\alpha_1 \right) \nonumber  \\
&& = \frac{1}{N_2}\left( N_2-N_1 + \frac{(M-N_2)^2\alpha_1\alpha_2}\Delta - (M-N_1)\alpha_1 \right) \nonumber \\
&& \overset{(a)}{\le} 0,  
\end{eqnarray}
where $(a)$ is due to $ \frac{(M-N_2)^2\alpha_1\alpha_2}\Delta \le (M-N_1)\alpha_1 - (N_2-N_1)$. Otherwise, if $N_2-N_1 + \frac{(M-N_2)^2\alpha_1\alpha_2}\Delta > (M-N_1)\alpha_1$, it can be checked by the same steps that $\mathcal{P}_{123}$ violates $\ell_4$ and does not exist. 
corner point with intersection of $\ell_1$, $\ell_2$, and $\ell_4$ is given by $\mathcal{P}_{124} = ((M-N_1)\alpha_1 - (N_2 - N_1), (M-N_1)\alpha_1, N_2 - (M-N_1)\alpha_1)$, which requires $\frac{N_2-N_1}{M-N_1} \le \alpha_1$ for non-negativity of $d_1$ and $\alpha_1 \le \frac{N_2}{M-N_1}$ for non-negativity of $d_0$. If $ (M-N_1)\alpha_1 - (N_2-N_1) \le \frac{(M-N_2)^2\alpha_1\alpha_2}\Delta$, $\mathcal{P}_{124}$ satisfies $\ell_3$, since $
d_1 + d_2 + d_0 = N_1 + (M-N_1)\alpha_1$, 
then
\begin{eqnarray}
&& \!\!\!\!\!\!\!\!\!\!\! d_1 + d_2 + d_0 - \left( N_2+\frac{\alpha_{1}\alpha_{2}(M-N_2)^2}{(N_2-N_1)(1-\alpha_{1})+(M-N_2)\alpha_{2}}\right) \nonumber \\
&& \!\!\!\!\!\!\!\!\!\!\! = N_1 - N_2 + (M-N_1)\alpha_1 - \nonumber \\
&& \!\!\!\!\!\!\!\!\!\!\!  \quad \frac{\alpha_{1}\alpha_{2}(M-N_2)^2}{(N_2-N_1)(1-\alpha_{1})+(M-N_2)\alpha_{2}}    \overset{(a)}{\le} 0,
\end{eqnarray}
where $(a)$ is due to $ (M-N_1)\alpha_1 - (N_2-N_1) \le \frac{(M-N_2)^2\alpha_1\alpha_2}\Delta$. If $  \frac{(M-N_2)^2\alpha_1\alpha_2}\Delta$ $< (M-N_1)\alpha_1 - (N_2-N_1)$, it can be checked by the same steps that $\mathcal{P}_{124}$ violates $\ell_3$ and does not exist. The corner point with intersection $\ell_1, \ell_3$, and $\ell_4$ does not exist, because the determinant of coefficient matrix for corresponding equations is zero.

\end{appendices}

\bibliographystyle{IEEEtran}
\bibliography{RSMA}

\end{document}